\documentclass[twocolumn,prb,showpacs,preprintnumbers,amsmath,amssymb]{revtex4}

\usepackage{graphicx}% Include figure files
\usepackage{psfig}% Include figure files
\usepackage{dcolumn}% Align table columns on decimal point
\usepackage{bm}% bold math

\begin{document}

% \preprint{AFM$_d$/chiAFM1$_d$/AF17}

% \renewcommand{\thefootnote}{\fnsymbol{footnote}}

\title{Magnetic structures
and reorientation transitions in noncentrosymmetric uniaxial
antiferromagnets}

\author{A.N.\ Bogdanov}
\altaffiliation[Permanent address: ]%
{Donetsk Institute for Physics and Technology,
340114 Donetsk, Ukraine;  
}
\email{bogdanov@kinetic.ac.donetsk.ua}
\author{U.K. R\"o\ss ler}
\thanks
{Corresponding author: IFW Dresden,
Postfach 270116, D--01171 Dresden, Germany. 
Tel.: +49-351-4659-542; Fax: +49-351-4659-537}
\email{u.roessler@ifw-dresden.de}
\author{M. Wolf}
\author{K.-H. M\"uller}
\affiliation{
Leibniz-Institut f{\"u}r Festk{\"o}rper- 
und Werkstoffforschung Dresden\\
Postfach 270116
D--01171 Dresden, Germany
}%

\date{\today}

\begin{abstract}
A phenomenological theory of magnetic states 
in noncentrosymmetric tetragonal 
antiferromagnets is developed,
which has to include homogeneous and inhomogeneous 
terms (Lifshitz-invariants) 
derived from Dzyaloshinskii-Moriya couplings.
Magnetic properties of this class of antiferromagnets 
with low crystal symmetry are discussed in relation 
to its first known members,
the recently detected compounds
Ba$_2$CuGe$_2$O$_7$ 
and  K$_2$V$_3$O$_8$.
Crystallographic symmetry
and magnetic ordering in these systems 
allow the simultaneous occurrence 
of chiral inhomogeneous magnetic structures 
and weak ferromagnetism.
New types of incommensurate
magnetic structures are possible, namely, 
{\textit {chiral helices}} with rotation of 
staggered magnetization and oscillations of 
the total magnetization.
Field-induced reorientation transitions 
into modulated states have been studied 
and corresponding phase diagrams are constructed.
Structures of magnetic defects (domain-walls 
and vortices) are discussed. 
In particular, vortices, i.e. localized non-singular 
line defects, are stabilized 
by the inhomogeneous Dzyaloshinskii-Moriya interactions 
in uniaxial noncentrosymmetric antiferromagnets.
\end{abstract}

\pacs{
75.10.-b,
75.50.Ee,
75.30.Kz
%
% 75.25.+z Spin arrangements in magnetically ordered materials 
% (including neutron and spin-polarized electron studies, 
% synchrotron-source x-ray scattering, etc.)
% 75.10.-b General theory and models of magnetic ordering 
% (see also 05.50 Lattice theory and
% statistics)
% 75.10.Hk Classical spin models
% 75.30.Gw Magnetic anisotropy
% 75.30.Kz Magnetic phase boundaries (including magnetic transitions, 
% metamagnetism, etc.)
% 75.30.Et Exchange and superexchange interactions 
% (see also 71.70 Level splitting and interactions)
% 75.50.Ee Antiferromagnetics
}
% %%% PACS numbers

\maketitle

% \clearpage

\section{Introduction}\label{intro}

In many magnetic crystals the magnetic properties
are strongly influenced by the antisymmetric exchange
({\textit {Dzyaloshinskii-Moriya}}) coupling
which is generally described by 
a vector product formed by the magnetic moments ${\mathbf S}_i$ 
of two magnetic ions:
\begin{equation}
w_D={\mathbf D}_{ij}\cdot({\mathbf S}_{i}\times {\mathbf S}_{j})
\label{Dz}
\end{equation}
and 
the so-called Dzyaloshinskii vector ${\mathbf D}_{ij}$.\cite{Dz57,Moriya}
Based on phenomenological considerations,
the interaction (\ref{Dz}) was introduced by Dzyaloshinskii 
to explain the observation of a small net magnetization
in a number of antiferromagnets, 
a phenomenon called {\textit {weak ferromagnetism}}\cite{Neel53}
which is due to a slight deviation of 
the sublattice magnetizations from antiparallel arrangement.
Extending Anderson's theory of superexchange 
Moriya later found 
a microscopic mechanism 
due to spin--orbit interactions
that is responsible for the interactions (\ref{Dz}).
They arise in certain groups of magnetic crystals 
with low symmetry 
where the effects the couplings (\ref{Dz}) 
do not cancel.\cite{Moriya}
During the following decades intensive theoretical and 
experimental studies 
on the {\textit {Dzyaloshinskii-Moriya}} coupling (\ref{Dz}) 
resulted in a deep insight into its microscopic origins 
and its manifestation in macroscopic properties of 
magnetic materials.\cite{Chikazumi,Sandratskii98}
Now it is known that weak ferromagnetism must essentially 
be attributed to many types of complex magnetic structures.
It influences appreciably the magnetic properties 
of several important classes of magnetic materials 
such as orthoferrites, manganites, 
some high-temperature superconducting cuprates,
and others.\cite{Chikazumi,Luo99}

Another fundamental macroscopic manifestation of
antisymmetric couplings (\ref{Dz}) takes place 
in noncentrosymmetric magnetic crystals.
Dzyaloshinskii showed that,
in this case, 
the interaction (\ref{Dz})
stabilizes long-periodic spatially modulated
structures with {\textit {fixed}} sense of rotation of the
vectors ${\mathbf S}_i$.
\cite{Dz64}
Within a continuum approximation for magnetic properties,
the interactions responsible for these modulations 
are expressed by inhomogeneous invariants. 
We will call these contributions 
to the (free) magnetic energy,
involving first derivatives of magnetization 
or staggered magnetization with respect
to spatial coordinates,
{\textit {inhomogeneous}} Dzyaloshinskii-Moriya interactions.
They are linear with respect to the first 
spatial derivatives of the magnetization ${\mathbf M}$ of type
\cite{Dz64}
\begin{equation}
 M_i\frac{\partial M_j}{\partial \eta}-M_j\frac{\partial M_i}
{\partial \eta}  \, , \label{lifshitz}
\end{equation}
where $M_i$, $M_j$ are components of magnetization 
vector(s) that arise in certain combinations 
in expressions (\ref{lifshitz}) depending on crystal symmetry, 
and ${\eta}$ is a spatial coordinate.\cite{Dz64} 
Such antisymmetric mathematical forms 
first have been studied 
in the theory of phase transitions 
by E. M. Lifshitz \cite{Landau5} 
and are known as {\textit {Lifshitz invariants}}.
The first magnetic modulated chiral structures 
predicted in \cite{Dz64} were observed 
in the cubic noncentrosymmetric crystals MnSi 
and FeGe (space group $P2_13$).\cite{MnSi,Bak80}
During the following years 
modulated magnetic structures of this kind 
were discovered and investigated in several 
classes of magnetic crystals 
lacking inversion center.
\cite{Izyumov84,Lebech89,Adachi80}
These {\textit {chiral}} helical structures 
are essentially different from
numerous other spatially modulated magnetic states 
in systems with competing exchange interactions 
(as, e.g., in rare earth metals).
\cite{Izyumov84,Koehler65}
The latter are characterized by rather short
periods (usually including only few unit cells) 
and arbitrary rotation sense. 
On the contrary, chiral structures due 
to (\ref{Dz}) or (\ref{lifshitz})
have long period and a fixed sense of rotation. 
For example, in MnSi the periodicity lengths of the helix 
in zero magnetic field was found 
to be about 170~{\AA} (39~unit cells), 
and FeGe has an even larger period (700~{\AA} 
or 149~unit cells).\cite{MnSi}
The interactions of type (\ref{lifshitz}) may also stabilize
periodic structures modulated 
in two-dimensions ({\textit vortex lattices})
and localized axisymmetric inhomogeneous 
states.\cite{JETP89}

Up to now both physical effects induced 
by the antisymmetric exchange coupling (\ref{Dz}) ---
{\textit {weak-ferromagnetism}} and {\textit {chiral modulations}} ---
never have been observed simultaneously 
in one magnetic system. 
Moreover, in noncentrosymmetric magnetic crystals 
with chiral modulations that were known and described
so far, the existence of weak ferromagnetism
is excluded because of their symmetry. 
\cite{Izyumov84}
In this paper, we show that both phenomena
can coexist in the recently discovered
noncentrosymmetric tetragonal antiferromagnets 
Ba$_2$CuGe$_2$O$_7$ \cite{Zhelud96,Zhelud97}
and  
K$_2$V$_3$O$_8$.\cite{Lumsden}
Due to the crystallographic and magnetic structures 
of these crystals, the Dzyaloshinskii-Moriya 
coupling (\ref{Dz}) favours noncollinearity 
along one direction and spatial modulations along the others.
Here we determine possible magnetic phases 
and study their evolution in applied magnetic fields.
It turns out that 
the unique coexistence of weak ferromagnetism
and chiral modulations enables the occurrence of new
types of incommensurate structures and specific localized structures
including so called {\textit {magnetic vortices}} or {\textit {skyrmions}}
which are generally unstable in other classes of magnetic materials.

\section{The model}\label{model}

\subsection{Phenomenological energy}\label{Energy}
The tetragonal antiferromagnet Ba$_2$CuGe$_2$O$_7$ 
\cite{Zhelud96,Zhelud97} (space group $P\bar{4}2_{1}m$)
belongs to the crystallographic class 
$D_{2d}$
and K$_2$V$_3$O$_8$ \cite{Lumsden} (space group $P4bm$) 
to $C_{4v}$. 
The magnetic (free) energy within a continuum description 
consistent with the symmetry 
and the two-sublattice magnetic structure of 
these antiferromagnets can be derived by 
the standard approach to phenomenological theory.\cite{Landau5}
At temperatures sufficiently below the ordering temperature
the vectors of sublattice magnetization ${\mathbf M}_i$ ($i=1,2$)
do not change their modulus. 
In this practically important 
case defined by neglecting the {\textit {paraprocess}}, 
the vectors ${\mathbf M}_i$ have only orientational degrees
of freedom and can be described by the unity vectors 
${{\mathbf m}_i={\mathbf M}_i/M_s}$, 
where ${{M_s}=|{\mathbf M}_1|=|{\mathbf M}_2|}$ is the sublattice 
saturation magnetization.
For tetragonal antiferromagnetic crystals,
the two-sublattice model
described by the unity vectors ${\mathbf m}_1$ and 
${\mathbf m}_2$
yields the magnetic energy 
in the following form:
\begin{eqnarray}
W&=\int\Bigg\{&\frac{{\alpha}}{2}
\sum_{i=1}^{3}\left[\left(\frac{\partial{\mathbf m}_1}
{\partial x_i}\right)^2+\left(\frac{\partial{\mathbf m}_2}
{\partial x_i}\right)^2\right] 
\nonumber\\
& & + {\alpha}'\sum_{i=1}^{3}\left(\frac{\partial{\mathbf m}_1}
{\partial x_i}\frac{\partial{\mathbf m}_2}{\partial x_i}\right)+ 
\frac{{\lambda}}{2}{\mathbf m}_1\cdot{\mathbf m}_2 
\nonumber\\
& & -{\mathbf h}\cdot({\mathbf m}_1 + {\mathbf m}_2) 
\nonumber\\
& & 
-\frac{{\beta}}{2}(m_{1z}^2 + m_{2z}^2) - {\beta}'m_{1z}m_{2z} 
\nonumber\\
& & 
-d\,(m_{1x}m_{2y}-m_{2x}m_{1y}) + w_D \Bigg\}\,d\,V.
\label{energy0}
\end{eqnarray}
This includes inhomogeneous (${\alpha}$, ${\alpha}'$) 
and homogeneous (${\lambda }$) parts of the exchange
coupling and the interaction energy 
with the external field ${\mathbf h}$. 
The next two terms describe 
uniaxial second-order magnetocrystalline anisotropy 
with constants ${\beta}$, ${\beta}'$,
where the $z$-axis is taken along 
the tetragonal axis of the antiferromagnets. 
The homogeneous part of the Dzyaloshinskii-Moriya interaction 
with constant $d$ is responsible for weak ferromagnetism
with small magnetic moments in the basal plane.
Finally the energy contribution $w_D$ includes 
Lifshitz invariants of type (\ref{lifshitz}). 
The functional form of $w_D$
depends on the crystal symmetry and will be specified later. 

The next terms in a systematic expansion of the energy 
for a two-sublattice antiferromagnet are
much weaker fourth-order terms of the magnetocrystalline
anisotropy, including uniaxial parts with terms  
${m_{1z}^4}$, ${m_{2z}^4}$, 
${m_{1z}^2m_{2z}^2}$,
and a magnetic anisotropy in the basal plane ($XOY$ plane) 
composed of ${x}$-- and ${y}$--components of the vectors ${\mathbf m}_i$.
The former are important in close vicinity to some reorientation
transitions and the latter is responsible for small variations
of magnetic structures 
when the magnetic field is rotated in the basal plane.
These secondary effects are omitted in this contribution dedicated 
to the principal magnetic properties of the system.
We also neglect the stray field contribution 
in the total energy (\ref{energy0})
because, due to the antiparallel alignment of
magnetic moments in antiferromagnets, 
stray fields are much weaker than in ferromagnetic crystals. 
They, however, play a crucial role 
in stabilizing multidomain structures 
in the vicinity of field-induced 
reorientation transitions.\cite{Bar88}

The functional (\ref{energy0}) includes 
all leading interactions
in an uniaxial two-sublattice
antiferromagnetic crystal.
Here, we briefly list several special cases of the model
(\ref{energy0}) which describe
important special classes of antiferromagnetic systems.\\
I.) $d$ = 0, $w_D$ = 0.  {\textit {Collinear antiferromagnets}}.
The vast group of these antiferromagnetic materials 
includes such well-studied species
as CuCl${_2 \cdot}$2H$_2$O \cite{Lowe72}, MnF$_2$ \cite{Shapira70}, 
Cr$_2$O$_3$ \cite{Shapira77}, GdAlO$_3$ \cite{Rohrer68} 
(see for further references and 
review of their magnetic properties
Refs.~\cite{Bar88} and \cite{Dejongh74}).\\
II.) $d \ne $ 0,  $w_D$ = 0. 
{\textit {Antiferromagnets with weak ferromagnetism}}.
In this case the energy (\ref{energy0}) describes  
antiferromagnetic crystals 
with homogeneous Dzyaloshinskii-Moriya interactions 
resulting in weak ferromagnetism 
or collinear antiferromagnets with
``hidden'' weak ferromagnetism. 
Among many others, 
this group includes MnCO$_3$,\cite{Neel53}
orthoferrites,\cite{Belov}
manganites,\cite{WFM in manganites}
and the most popular and 
well-studied weak-ferromagnetic antiferromagnet  
{\textit {hematite}}, 
i.e. ${\alpha}$-Fe$_2$O$_3$.\cite{Dz57,Moriya,hematite,FNT86,Morrish94}
\\
III.) $d$ = 0, $w_D \ne $ 0. {\textit {Chiral helimagnets}}.
This case is realized 
in the cubic helimagnets discussed above 
and in other noncentrosymmetric magnetic systems.
\cite{MnSi,Izyumov84} 
Usually interactions of type (\ref{lifshitz}) stabilize 
modulated chiral structures in these materials \cite{Izyumov84}.\\
IV.) $d \ne $ 0, $w_D \ne $ 0. 
{\textit {Chiral helimagnets with weak ferromagnetism}}.
The model (\ref{energy0}) represents previously unknown systems 
where both, homogeneous and inhomogeneous, Dzyaloshinskii-Moriya 
interactions are operational and not forbidden by any other
additional symmetries.
As will be shown in this paper, this unique coexistence 
of a mechanism for weak ferromagnetism and chiral coupling 
leads to specific modulated states with magnetization oscillations.
Thus, in the following we will generally consider systems
$d \ne 0$ and $w_D \ne 0$.

\subsection{Simplified model and basic equations}\label{simplemodel}
It is convenient to use linear combinations of 
the sublattice magnetization 
${\mathbf m}_i$, namely, the vector of {\textit total} magnetization 
${\mathbf m}=({\mathbf m}_1+{\mathbf m}_2$)/2 
and the {\textit {staggered}} magnetization 
(or vector of {\textit {antiferromagnetic}} order)
${\mathbf l}=({\mathbf m}_1-{\mathbf m}_2$)/2
as internal variables of the system.
Because of $|{\mathbf m}_{i}|$ = 1
these vectors satisfy the constraints 
${\mathbf m}\cdot{\mathbf l}$ = 0, ${\mathbf m}^2+{\mathbf l}^2 = 1$.

In most antiferromagnetic crystals the exchange coupling is
much stronger than other internal interactions.
Strong magnetic fields of order ${\lambda}$ 
destroy antiferromagnetic order and orientate 
the sublattice magnetizations ${\mathbf m}_{i}$ parallel 
to each other (a so called {\textit spin-flip} transition 
into the ``paramagnetic''
phase with $|{\mathbf m}|$ = 1, ${\mathbf l}$ = 0). 
For most investigated antiferromagnetic systems 
these ``exchange'' fields
are extremely large. Practically attainable 
values of magnetic fields
usually only slightly distort 
the antiparallel arrangement 
inducing states with 
the total magnetization much smaller than unity. 
The hierarchy for the strength of interactions,
${\lambda}\gg d, {\beta}, {\beta'}$,
and the relations for the internal parameters,
${m \ll }$ 1 and ${l \approx }$ 1, 
permit
to considerably simplify the energy (\ref{energy0}) 
by excluding gradients of ${\mathbf m}$ and taking into 
account only the following terms
(for details, see e.g. \cite{FNT86})
\begin{eqnarray}
\widetilde{W} & = %
\int \Bigg\{&A\sum_{i=1}^{3}%
\left(\frac{\partial{\mathbf l}}{\partial x_i}\right)^2
+{\lambda }{\mathbf m}^2-2{\mathbf m}\cdot{\mathbf h}
\nonumber \\
& & 
+ 2d\,(m_xl_y-m_yl_x)-Bl_z^2+w_D \Bigg\}\,d\,V\,,
\label{energy}
\end{eqnarray}
where $A={\alpha}-{\alpha}'$ and $B={\beta} - {\beta}'$.
Functional forms of $w_D$ for all noncentrosymmetric
crystallographic classes have been derived in Ref.~\onlinecite{AFM89}. 
In particular, for the antiferromagnets under consideration 
the Lifshitz invariants
quadratic in the components of ${\mathbf l}$ have the following form
\begin{equation}
{\textrm {class\ }}{\textit D}_{2d}\,{\textrm {:}} \;
w_D=D\left( l_z\frac{\partial l_x}{\partial y}-l_x\frac{\partial l_z}{\partial
y}+l_z\frac{\partial l_y}{\partial x}-l_y\frac{\partial l_z}{\partial x}%
\right)\, ,
\label{lifshitzC}
\end{equation}
\begin{equation}
{\textrm {class\ }}{\textit C}_{nv}\,{\textrm {:}} \;
w_D=D\left( l_z\frac{\partial l_x}{\partial x}-l_x\frac{\partial l_z}{\partial
x}+l_z\frac{\partial l_y}{\partial y}-l_y\frac{\partial l_z}{\partial y}%
\right)\,.
\label{lifshitzD}
\end{equation}
The homogeneous part of 
the Dzyaloshinskii-Moriya interaction 
in (\ref{energy}) includes
in-plane components of ${\mathbf m}$ and ${\mathbf l}$.
They originate from the $z$-component
of the vector product (\ref{Dz}). 
Writing the Dzyaloshinskii vector
in (\ref{energy}) as a sum of two parts
proportional to $D$ and $d$,
this contribution to (\ref{energy})
can be derived from the vector directed 
along the tetragonal axis ${\mathbf d}= (0, 0, d)$. 
On the other hand, the Lifshitz invariants
(\ref{lifshitzC}), (\ref{lifshitzD})
can be derived by an expansion for the in-plane components
of the vector product (\ref{Dz})
considering the contribution 
due to the vector ${\mathbf D}= (D, D, 0)$.
The terminology in this field is not yet fixedly formulated. 
Here, following Ref.~\onlinecite{JMMM94}, we will
call energy contributions given by Lifshitz invariants 
{\textit {Dzyaloshinskii}} or {\textit {chiral}} interactions 
to distinguish them from 
the homogeneous part of the Dzyaloshinskii-Moriya interaction.

Independent minimization of the energy (\ref{energy})
with respect to ${\mathbf m}$ leads to the following result 
\begin{equation}
{\mathbf m}=%
-[{\mathbf n}\times({\mathbf d}+{\mathbf n}\times{\mathbf h})]/{\lambda} \,,
\label{m}
\end{equation}
where ${\mathbf n}={\mathbf l}/|{\mathbf l}|$ is the unity vector parallel 
to the staggered magnetization vector.
After substitution of (\ref{m}) 
into the energy (\ref{energy}) 
one obtains the energy 
to leading approximation 
as a function of the vector ${\mathbf n}$
\begin{eqnarray}
\widetilde{W} & = %
\int \Bigg\{&A\sum_{i,j}\left(\frac{\partial {n}_{i}}
{\partial x_{j}}\right)^2
\nonumber\\
& & -\frac{1}{{\lambda}}[(h_x+d{n}_y)^2+(h_y-d{n}_x)^2-
({\mathbf h}\cdot{\mathbf n})^2]
\nonumber\\
& & -B{n}_z^2+w_D ({\mathbf n})\Bigg\}\,d\,V
\,,
\label{energy2}
\end{eqnarray}
where $w_D ({\mathbf n})$ is determined 
by (\ref{lifshitzC}) or (\ref{lifshitzD}).
The energy (\ref{energy2}) and (\ref{m}) were 
derived from (\ref{energy0})
by ignoring the paraprocess and assuming weak total magnetization,
${|{\mathbf m}| \ll 1}$, 
implying ${|{\mathbf l}| \simeq 1}$.
Both assumptions are fulfilled 
in most realistic cases of interest.
Thus, the energy (\ref{energy2}) describing
the orientation of the staggered magnetization 
can be considered as general phenomenological 
description for realistic uniaxial two-sublattice antiferromagnets.
The functional (\ref{energy2}) is related to so--called
nonlinear ${\sigma}$-models which are basic subjects
in the theory of solitons and which are intensively 
studied in mathematical and theoretical physics.\cite{Rebbi}

The energy contributions (\ref{lifshitzC}) and (\ref{lifshitzD})
can result in states with modulations 
in the basal tetragonal plane, 
i.e. with a propagation vector in the $XOY$-plane.
Its actual direction is selected 
by some small in-plane anisotropy contributions 
that are neglected here 
(see discussion on expansion (\ref{energy0}) above).
On the other hand, there are no interactions in our systems 
violating homogeneity along the tetragonal $z$-axis.
Hence, we infer that within the model (\ref{energy2})
the most general solutions are inhomogeneous 
only in the basal plane 
but homogeneous along the $z$-axis.
It is convenient to write the vector ${\mathbf n}(x,y)$
and the magnetic field ${\mathbf h}$ 
in spherical coordinates:
\begin{eqnarray}
{\mathbf n} & = &
  (\sin{\theta}\cos{\psi}, \sin{\theta}\sin{\psi}, \cos{\theta}),
\nonumber \\
{\mathbf h} & = &
 (h\sin{\zeta}\cos{\eta}, h\sin{\zeta}\sin{\eta}, h\cos{\zeta}) \,.
\label{spherical}
\end{eqnarray}
In these variables the total energy is given by
\begin{eqnarray}
\widetilde{W} &  = L_z 
\int\Bigg\{&A\sum_{i=1}^{2}\left[\left(\frac{\partial{\theta}}
{\partial x_i}\right)^2+
\sin^2{\theta}\left(\frac{\partial{\psi}}
{\partial x_i}\right)^2\right]
\nonumber\\
& & +w_D +\tilde{w} \Bigg\}\,dx\,dy\,,
\label{energyT}
\end{eqnarray}
where the integration with respect to $z$ was performed 
for a system with linear size $L_z$, $x_1=x$, $x_2=y$.
The Lifshitz invariants are given by
\begin{eqnarray}
w_D= & D\Bigg[&\sin{\psi}\frac{\partial{\theta}}{\partial x}
+ \cos{\psi}\frac{\partial{\theta}}{\partial y}
\label{lifshitzD3} 
\\
& &+\sin{\theta}\cos{\theta}\left(\cos{\psi}
\frac{\partial{\psi}}{\partial x}
- \sin{\psi}\frac{\partial{\psi}}{\partial y}\right)\Bigg]
\;{\textrm {for}}\;{\textit D}_{2d}\,,
\nonumber
\end{eqnarray}
\begin{eqnarray}
w_D= & D\Bigg[&\cos{\psi}\frac{\partial{\theta}}{\partial x}
+\sin{\psi}\frac{\partial{\theta}}{\partial y}
\label{lifshitzC3}
\\
& &-\sin{\theta}\cos{\theta}\left(\sin{\psi}
\frac{\partial{\psi}}{\partial x}
- \cos{\psi}\frac{\partial{\psi}}{\partial y}
\right)\Bigg] \; {\textrm {for}}\; {\textit C}_{nv} \,,
\nonumber
\end{eqnarray}
and the energy term ${\tilde{w}}$ 
does not depend on spatial derivatives:
\begin{eqnarray}
{\lambda}\tilde{w}  & = &
-({\lambda}B-d^2-h^2\cos^2{\zeta})\cos^2{\theta}
\nonumber\\
& &
-(h^2\sin^2{\zeta}+d^2)
+h^2\sin^2{\zeta}\cos^2({\psi}-{\eta})\sin^2{\theta}
\nonumber\\
& &
-2dh\sin{\zeta}\sin{\theta}\sin({\psi}-{\eta}) 
\nonumber\\
& &
+h^2\sin{\zeta}\cos{\zeta}
\sin{2\theta}\cos({\psi}-{\eta})
\,.
\label{density0}
\end{eqnarray}
The functional (\ref{energyT}) 
with (\ref{density0}) provides 
the basic expression for the total energy 
of uniaxial two-sublattice antiferromagnets 
belonging to crystallographic 
classes without inversion symmetry.
By inserting the appropriate Lifshitz invariant from
(\ref{lifshitzD3}) or (\ref{lifshitzC3}) for $w_D$,
the functional describes the magnetic energy 
of the two tetragonal crystals of interest here.

\section{The phase diagram of equilibrium solutions}\label{phase diagram}

The equilibrium distributions of ${\theta}(x,y)$ and
${\psi}(x,y)$ are determined by solving 
a set of equations minimizing 
the energy (\ref{energyT}). 
% and (\ref{density0}).
Depending on the values of the phenomenological constants 
in the energy (\ref{energyT})
and the components of magnetic fields 
different spatially homogeneous and modulated
phases can be realized in the system.
Due to isotropy of the model in the basal plane 
only the component of magnetic field along the tetragonal axis
($h_z$) and the value of its projection onto the basal plane ${h_{\perp}}$ 
are of importance.  
% A reduction of the number of control parameters
% can be introduced by rescaling of the spatial variables 
% and the energy.
% It is convenient to rescale 
A reduction of the number of control parameters
is obtained by rescaling the spatial variables and the energy.
We use the following units 
for lengths, magnetic field, and the strength
$D$ of the inhomogeneous Dzyaloshinskii-Moriya interaction 
\begin{eqnarray}
x_0=\sqrt{A{\lambda}/|K|}, & & \qquad h_0=\sqrt{|K|},
\nonumber\\
D_0=\frac{4}{{\pi}}\sqrt{A|K|/{\lambda}}, & & 
\qquad K= {\lambda}B-d^2
\label{units}
\end{eqnarray}
introducing an effective anisotropy constant $K$
acting on the staggered magnetization
that is comparable to a constant of 
uniaxial anisotropy in ferromagnets.
In centrosymmetric antiferromagnets ($D=0$) 
and zero external field, 
the collinear state with staggered magnetization ${\mathbf l}$
along the tetragonal axis and $m=0$ 
is the ground state for $K>0$ ({\textit {easy-axis}} system); 
for $K<0$ the vectors ${\mathbf l}$ 
and a nonzero ${\mathbf m}$ 
lie in the in the basal plane 
(weak ferromagnetic states).
The characteristic length $x_0$ is of 
the order of the effective size of an
isolated domain wall 
between the homogeneous states at zero field.
In uniaxial ferromagnetic materials the corresponding expression 
for an intrinsic length is
known as {\textit {exchange}} or {\textit {Bloch}} length.\cite{Hubert98}
The characteristic field $h_0$ 
is the so--called {\textit {spin-flop}} field.
Finally, the  parameter $D_0$ equals  
the lowest value of the (Dzyaloshinskii) 
constant $D$ that stabilizes 
modulated states at zero field (see below).
In these reduced units (\ref{units}), 
the energy (\ref{energyT}) includes
as independent parameters the rescaled constants 
$K$ involving $d$ and $D$ 
as well as the two components of the applied field 
(${h_z}$, ${h_{\perp}}$).
Thus, these parameters span a four--dimensional 
phase--space for the solutions. 

% We start the analysis of the possible magnetic configuration
% in the system by pointing out some of their general features.
Before giving the detailed analysis, 
let us point out some general features of 
the possible magnetic configurations in this system.
The equilibrium magnetic structures are
governed by two opposing tendencies.
The rotation of the staggered vector ${\mathbf l}$ 
with propagation vectors in the basal plane 
and an appropriate sense of rotation leads 
to negative values of 
the invariants (\ref{lifshitzD3}), (\ref{lifshitzC3}). 
An unlimited reduction of the pitch for 
this winding of the staggered magnetizations
would lead to infinitely negative values of 
this Dzyaloshinskii energy.
This is counter-acted by the inhomogeneous part 
of the exchange energy in (\ref{energyT})
providing the ``stiffness'' of the magnetic structure.
In isotropic systems,
i.e. ${\tilde{w}}$=0 for expression (\ref{density0}),
the ratio of these competing energy contributions 
yields the optimal period for the spiral, 
which is of the order of $A/D$.\cite{Dz64}
Such chiral modulations with uniform rotation are observed
in low-anisotropy systems as cubic helimagnets \cite{Lebech89}
or in hexagonal chiral magnets with in-plane rotation
of the magnetization vectors.\cite{Adachi80}
The uniform rotation of ${\mathbf l}$ in spirals 
is disturbed by anisotropic interactions 
and/or by application of a magnetic field,
i.e. the energy terms included in (\ref{density0}).
These interaction terms result in
preferred directions 
for the staggered vector ${\mathbf l}$ 
corresponding to the minima 
of the energy density (\ref{density0}).
Hence, they distort the chiral modulations and
may even suppress them by forcing the staggered 
magnetization to point fixedly into ``easy'' directions. 
Thus, chiral modulations may occur only 
beyond a certain threshold: the interactions 
(\ref{lifshitzC}) or (\ref{lifshitzD}) 
must be strong enough to overcome 
the anisotropic energy contributions 
suppressing modulated states. 
Below this threshold the system takes 
on the homogeneous states which are determined 
by minimization of the energy (\ref{energyT}) with $D$ = $A$ = 0.
In the following subsections we will demonstrate this 
competition between homogeneous and inhomogeneous states
for our model in detail.

Another important property of the system is
related to the role played by the homogeneous Dzyaloshinskii-
Moriya interaction: in-plane components of the staggered
vector induce corresponding non-zero components of 
the total magnetization vector ${\mathbf m}$ 
(weak-ferromagnetic moments) according to eq.~(\ref{m}).
Thus, collinear antiferromagnetic states, ${\mathbf m}=0$,
exist only when the staggered vector ${\mathbf l}$ is parallel 
to the tetragonal axis.
All magnetic structures with a vector ${\mathbf l}$ 
deviating from this direction perforce have a locally 
nonzero magnetization ${\mathbf m}$.  
In zero field, these magnetic moments are in the basal plane. 
Helix-states of ${\mathbf l}$
are accompanied by magnetization components 
oscillating in sign and with the same period
as the antiferromagnetic modulations.
This peculiar mechanism leading 
to modulated antiferromagnetism 
and a related magnetization ${\mathbf m}$ 
may become operational and important, 
even if the ground-state is not modulated,
for the more general case ${\mathbf h}\neq0$ (see below).

In the general case the four dimensional phase diagram
(${d, D, h_z, h_{\perp}}$) includes regions 
with different modulated and homogeneous states separated by 
``hypersurfaces'' corresponding to different phase transitions.
% The known results for some specific case of our model give a useful
% information on general properties of the model.
As remarked above, there are numerous experimental 
and theoretical results on magnetic properties
of centrosymmetric antiferromagnets 
(model (\ref{energyT}) with $w_D=0$)
describing weak ferromagnetism.
\cite{Belov,hematite,FNT86}
The phase space of the control parameters in this case 
(${d, h_z, h_{\perp}}$)  was found to have a very
complex topology and, depending on the orientations
and the relative strengths of the vectors ${\mathbf h}$ and ${\mathbf d}$,
a number of non-trivial transitions
occur in these systems.\cite{FNT86}
This phase diagram can be considered
as a ``cross-section'' 
given by the three-dimensional ``hyperplane'' $D=0$
through the general (${d, D, h_z, h_{\perp}}$)-space
investigated here.
On the other side in chiral antiferromagnets 
described by the case $d$ = 0 for our model,
both the spin arrangements in a spiral and 
the corresponding propagation directions are found 
to be very sensitive to the orientation 
and strength of the applied field.\cite{FNT99}
In the general case of nonzero values of the constants 
$d$ and $D$, there is an even wider variety of 
homogeneous and inhomogeneous solutions characterized 
by complex noncollinear magnetic structures and variable 
directions of propagation vectors.
Clearly, the full set of the homogeneous and inhomogeneous
states corresponding to the energy (\ref{energyT}) 
together with (\ref{lifshitzD3}) to (\ref{density0})
are of general interest and well-worth 
of further investigations. 
Here, we restrict ourselves to the 
approximation valid for antiferromagnetic
systems with small magnetic anisotropy 
and their specific hierarchy of the interactions.
The physically expected relation 
$d \gg B$ turned out to be valid
in all known systems with weak ferromagnetism 
and is based on the common relativistic
origin of both magnetic energy contributions 
in these (low-anisotropy) systems.
It is specifically valid for systems
in which magnetism is due to d-electrons.
In these systems, the Dzyaloshinskii-Moriya interactions
generally overcome the magnetocrystalline anisotropy.
This is expected to
apply also in noncentrosymmtric antiferromagnets
with d-electron magnetism
for which the values for $d$ are still unknown. 
This allows to narrow considerably 
the range of physically meaningful control parameters
in our model.
\begin{figure*}
\includegraphics[width=8.0cm]{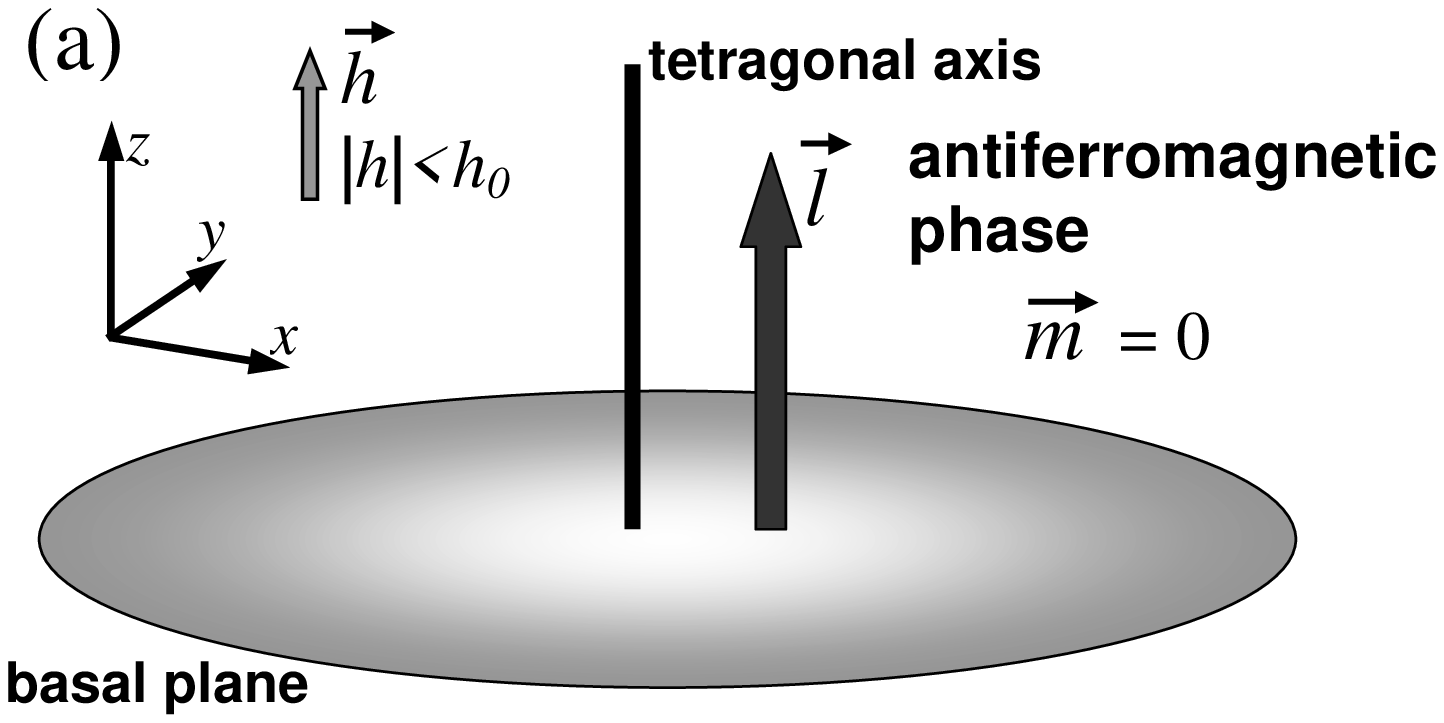}
\includegraphics[width=8.0cm]{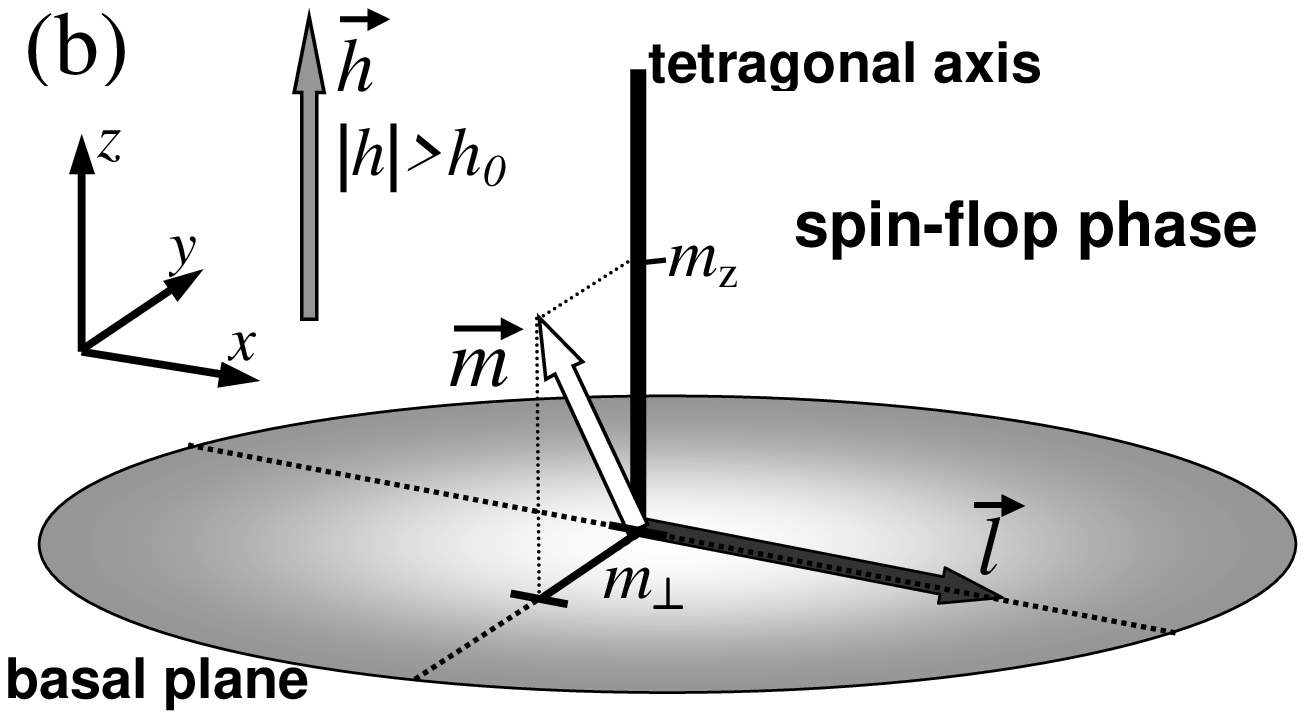}
\includegraphics[width=8.0cm]{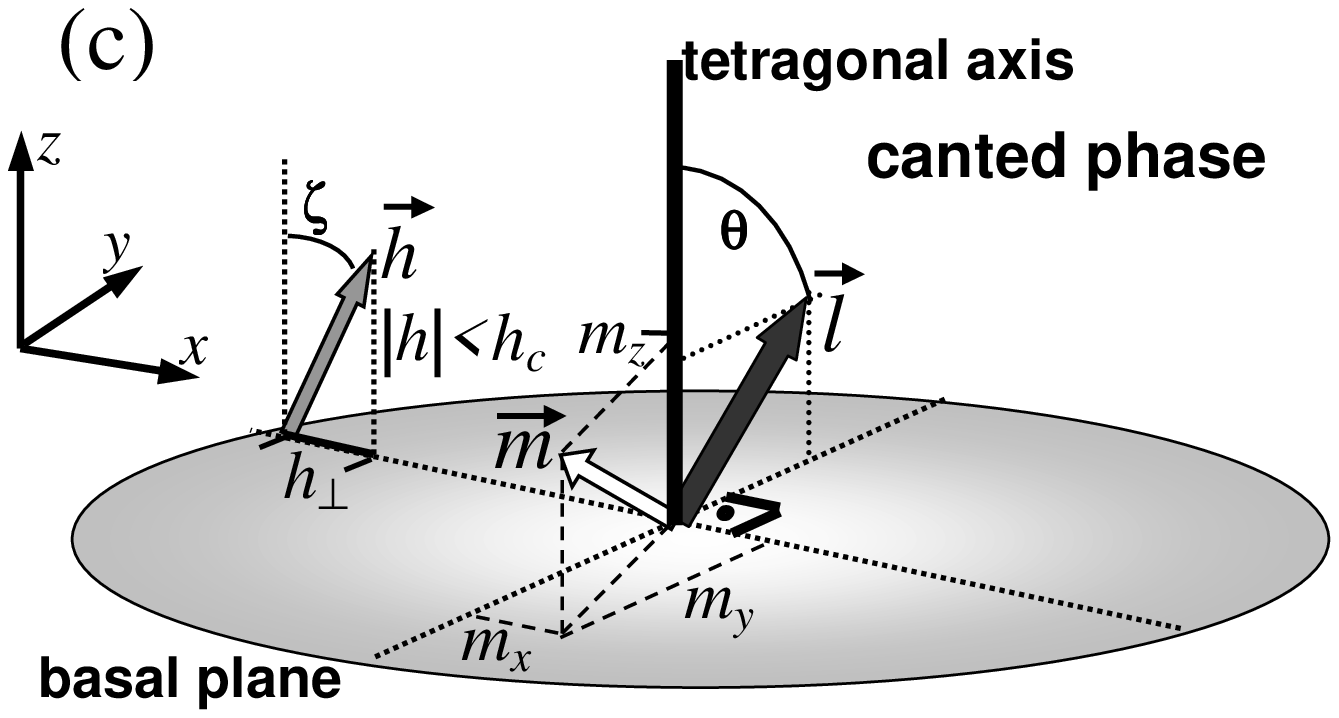}
\includegraphics[width=8.0cm]{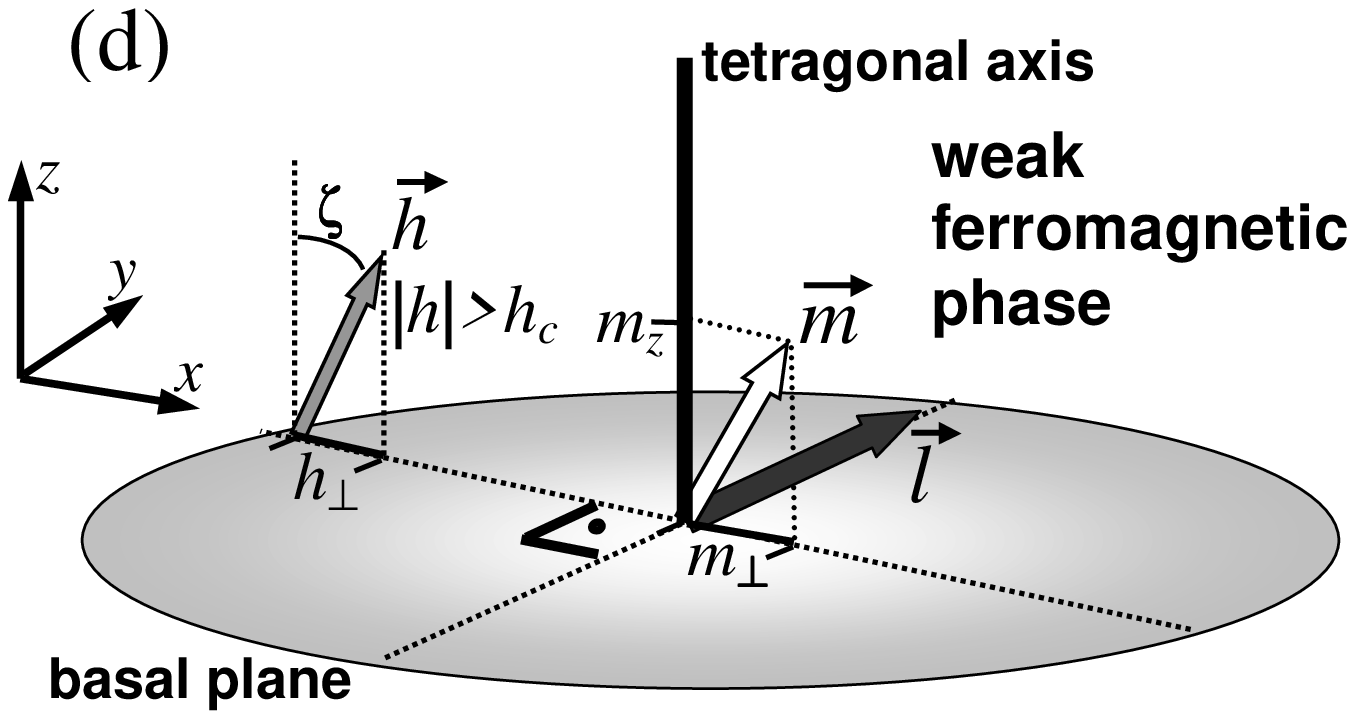}
\caption{
\label{fig1abcd}
Basic spin configurations 
described by the staggered magnetization ${\bf l}$
and the total magnetization ${\bf m}$ 
(for clarity, $|{\bf m}| \ll |{\bf l}| \simeq 1$ 
is not obeyed in the pictures) in homogeneous states
of easy-axis systems ($K>0$).\\
($a$) and ($b$) external field ${\bf h}$ along 
the tetragonal axis.
($a$) Collinear or {\textit {antiferromagnetic}} (AF) phase 
with ${\bf l} \parallel OZ$  and zero magnetization 
exists in the applied field along the tetragonal axis
${0 \le h \le h_0}$.\\
($b$) {\textit {Spin-flop}} phase with ${{\bf m} \parallel OZ}$ 
and with ${\bf l}$ in the basal plane is an 
important particular case of the {\textit {weak ferromagnetic}} (WF)
phase for ${h>h_0}$ along the tetragonal axis.\\
($c$) and ($d$)
magnetic field with oblique direction.
The low symmetry {\textit {canted}} phase ($c$)
exists for ${\nu}<1$ (\ref{density00})
and continuously transforms into the WF phase ($d$)
at the critical line  ${\nu}=1$. In the canted phase 
the staggered magnetization is in the plane perpendicular
to the the in-plane component of the applied field and
all components of the vector ${\bf m}$ have generally 
nonzero values (cf. eg.~(\ref{m2}).
($d$) General case of the WF phase. The staggered magnetization
${\bf l}$ lies in the basal plane, the component of the total
magnetization along the tetragonal axis (${m_z}$) 
is induced by the corresponding component of the applied field,
the in-plane components are due to the homogeneous
Dzyaloshinskii-Moriya interaction and in-plane components of magnetic 
field. In easy-plane systems ($K<0$) only the WF phase is stable
for all values of the applied field.
}
\end{figure*}

As already discussed above, 
the vector ${\mathbf d}$ induces 
a total magnetization in the basal plane 
which tends to orientate itself
parallel to the in-plane components of 
an applied field ${h_{\perp}}$.
Correspondingly, the staggered magnetization 
is rotated into the plane perpendicular 
the ${h_{\perp}}$ direction. 
The stronger the values
of $d$, ${h_{\perp}}$ and of 
the in-plane components of ${\mathbf l}$ 
the stronger is this effect.
We may assume that the deviation of the staggered magnetization 
from this plane perpendicular to the ${h_{\perp}}$ direction
${\varepsilon}=\left|{\psi}-{\eta}-{\pi}/2 \right|$
is small. 
By optimizing the energy (\ref{density0}) with respect
to ${\varepsilon}$ one obtains 
\begin{eqnarray}
{\varepsilon}=\left|{\psi}-{\eta}-{\pi}/2 \right|=
\left|\frac{h\cos{\zeta}\cos{\theta}}
{d+h\sin{\zeta}\sin{\theta}}\right|\,
\label{epsilon}
\end{eqnarray}
providing the consistency criterion 
for our assumption $\varepsilon \ll 1$.
This approximation is valid
always for $d\gg h$
and generally in a broad range of the orientations
for the vectors ${\mathbf h}$ and ${\mathbf l}$.
This includes almost all physically interesting cases. 
In the following analysis,
we assume as central approximation that the staggered magnetization
is always restricted to the plane perpendicular to ${h_{\perp}}$,
i.e. ${\varepsilon}=0$. 
By substituting ${{\psi}-{\eta}={\pi}/2}$ 
the energy density (\ref{density0}) can be simplified 
and reduced to the following form using the scaled 
quantities (\ref{units})
\begin{eqnarray}
{\tilde{w}} & = & \frac{|K|}{{\lambda}}\,{\Phi}({\theta}) 
\quad {\textrm{with}} 
\nonumber\\
& {\Phi}({\theta}) & =%
\textrm{sgn}K\left(1-\frac{h^2}{K}
\cos^2{\zeta}\right)\left(\sin{\theta}
-{\nu}\right)^2, \qquad 
\nonumber\\
& {\nu} & =%
\frac{dh\sin{\zeta}}{K-h^2\cos^2{\zeta}}\,.
\label{density00}
\end{eqnarray}
(Here, we drop constant terms in ${\tilde{w}}$,
i.e. those independent of $\theta$.)
In the following subsections, 
we investigate spatially homogeneous phases, 
helical phases, and their respective stability limits
with the approximation (\ref{density00}).
% Finally, localized chiral structures 
% (domain walls and/or kinks),
% arising as metastable defects 
% in these complex magnetic structures 
% are discussed.

\subsection{Homogeneous states}\label{Homogeneous States}

The homogeneous states are described by the
behaviour of the energy functional (\ref{density00}).
Depending on the sign of $K$ the energy (\ref{density00}) 
describes two different types of antiferromagnetic ordering.
\subsubsection{$K>0 \qquad$      Easy-axis system} \label{easyaxissys}

\begin{figure}
\includegraphics[width=8.0cm]{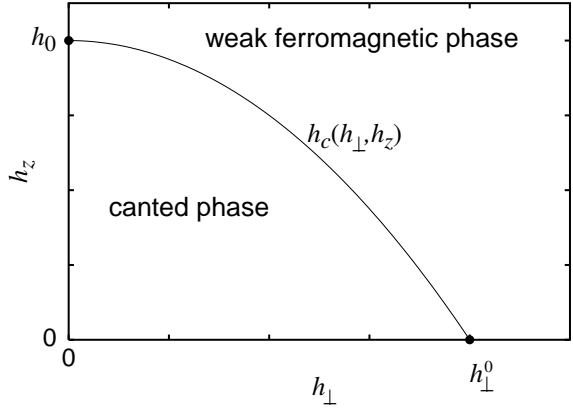}
\caption{
\label{fig2}
(${h_{\perp}, h_z}$) phase diagram 
for the homogeneous states in easy-axis systems ($K>0$).
The low symmetry canted phase 
transforms into the WF phase 
by a second order phase transition
at the critical line  $h_c$(${h_{\perp}, h_z}$)
determined by the condition ${\nu}=1$ in (\ref{density00}).
Scales are given by $h_0=K^{1/2}$ and $h_{\perp}^{0}=K/d$.
}
\end{figure}

At zero field and in a magnetic field 
along the tetragonal axis for  $h < h_0$ 
the 
{\textit {antiferromagnetic phase}}
with ${\mathbf l}\parallel$ $z$--direction 
and ${\mathbf m}$ = 0
has the lowest energy. 
This magnetic structure
is sketched in Fig.~\ref{fig1abcd}(a).
At the field ${h_0=\sqrt{K}}$ the vector ${\mathbf l}$
``flops'' down onto the basal plane. 
This is a so--called 
{\textit {spin-flop transition}}.
In the resulting {\textit {spin-flop phase}}
with ${\theta}$ = ${\pi}$/2,
the total magnetization 
under influence of the homogeneous Dzyaloshinskii-Moriya interaction 
is slightly inclined 
from the tetragonal axis (Fig.~\ref{fig1abcd}(b)). 
In the region where ${|{\mathbf m}| \ll 1}$
the components of ${\mathbf m}$,
${m_z=h/{\lambda}}$, ${|m_{\perp}|=d/{\lambda}}$,
are obtained from (\ref{m}).
In the region $ h_0<h <{\lambda}$ 
% increasing field 
the total magnetization increases linearly 
with increasing field
and finally at the ``exchange'' field 
$h_{\textrm{\footnotesize ex}}={\lambda}$ 
the spin--flop phase continuously
transforms into the saturated ``paramagnetic'' phase 
with 
${|{\mathbf m}|}$ = 1, ${\mathbf l}$ = 0
by a {\textit {spin-flip}} transition.
Note, in the spin-flop phase
the magnetic state has an infinite
degeneracy with respect to rotation of 
the vector ${\mathbf l}$ around the tetragonal axis. 
In--plane anisotropy reduces the degeneracy
to certain preferable directions related by symmetry 
in the basal plane. E.g., in the case of
fourth-order tetragonal anisotropy 
there are two mutually perpendicular
directions of ``easy'' magnetization.
In an increasing magnetic field deviating 
from the tetragonal axis
the staggered magnetization ${\mathbf l}$ rotates 
to the basal plane
in the plane perpendicular 
to the projection of ${\mathbf h}$
onto the basal plane.
The angle ${\theta}$ between 
the vector ${\mathbf l}$ and
the $z$-axis  is $\sin{\theta}={\nu}$ 
for ${\nu}<1$ (\ref{density00}).
We name this state with a finite angle 
between staggered magnetization and $z$-axis
{\textit {canted phase}}
(Fig.~\ref{fig1abcd}(c)).
Finally, at the critical line $h_c(h_{\perp},h_y)$
(Fig.~\ref{fig2}) where $\nu$ from expression (\ref{density00})
attains the critical value ${\nu}=1$,
a phase transition occurs into a phase 
with the staggered vector lying 
in the basal plane, $\sin{\theta}=1$, 
and perpendicular to the applied field (Fig.~\ref{fig1abcd}(d)).
This is a {\textit {weak ferromagnetic}} (WF) phase.
The phase-diagram for this transition 
between canted and WF phase is depicted in Fig.~\ref{fig2}.
The total magnetization is deduced 
by substituting 
the equilibrium values of ${\mathbf n}={\mathbf l}/|{\mathbf l}|$ 
into eq.~(\ref{m}).
Assuming that the applied field is 
in the $XOZ$-plane,
then the staggered magnetization rotates in the $YOZ$-plane.
From (\ref{m}) the
following expressions for the magnetization components result
\begin{eqnarray}
m_x & = & (h\sin{\zeta}+d\sin{\theta})/{\lambda},
\nonumber\\
m_y & = & - h\cos{\zeta}\sin{\theta}\cos{\theta}/{\lambda}, 
\nonumber\\
m_z & = & h\cos{\zeta}\sin^2{\theta}/{\lambda}\,,
\label{m2}
\end{eqnarray}
where ${\theta}=\arcsin{\nu}$ 
for the canted phase 
and
${\theta}={\pi}/2$ in the weak-ferromagnetic phase.

\begin{figure}
\includegraphics[width=8.0cm]{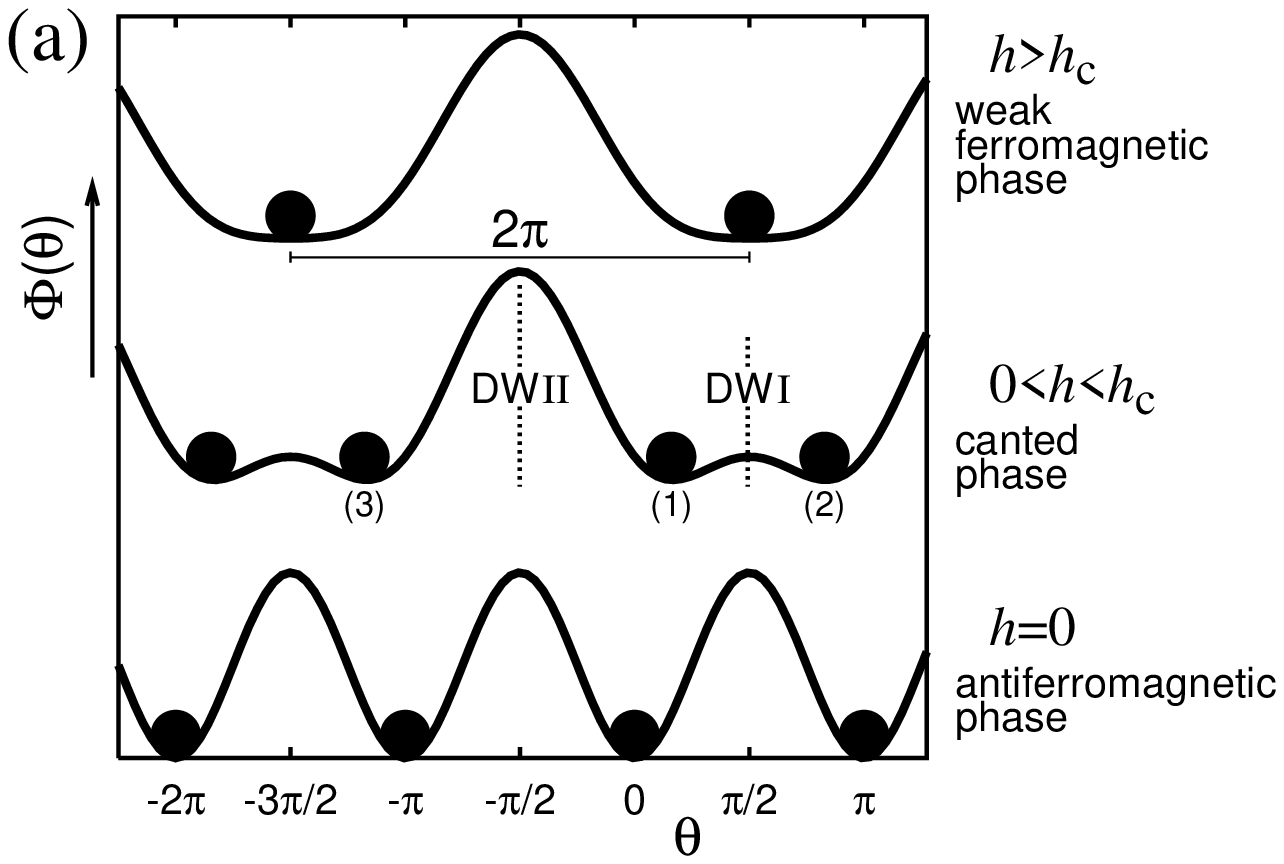}
\includegraphics[width=8.0cm]{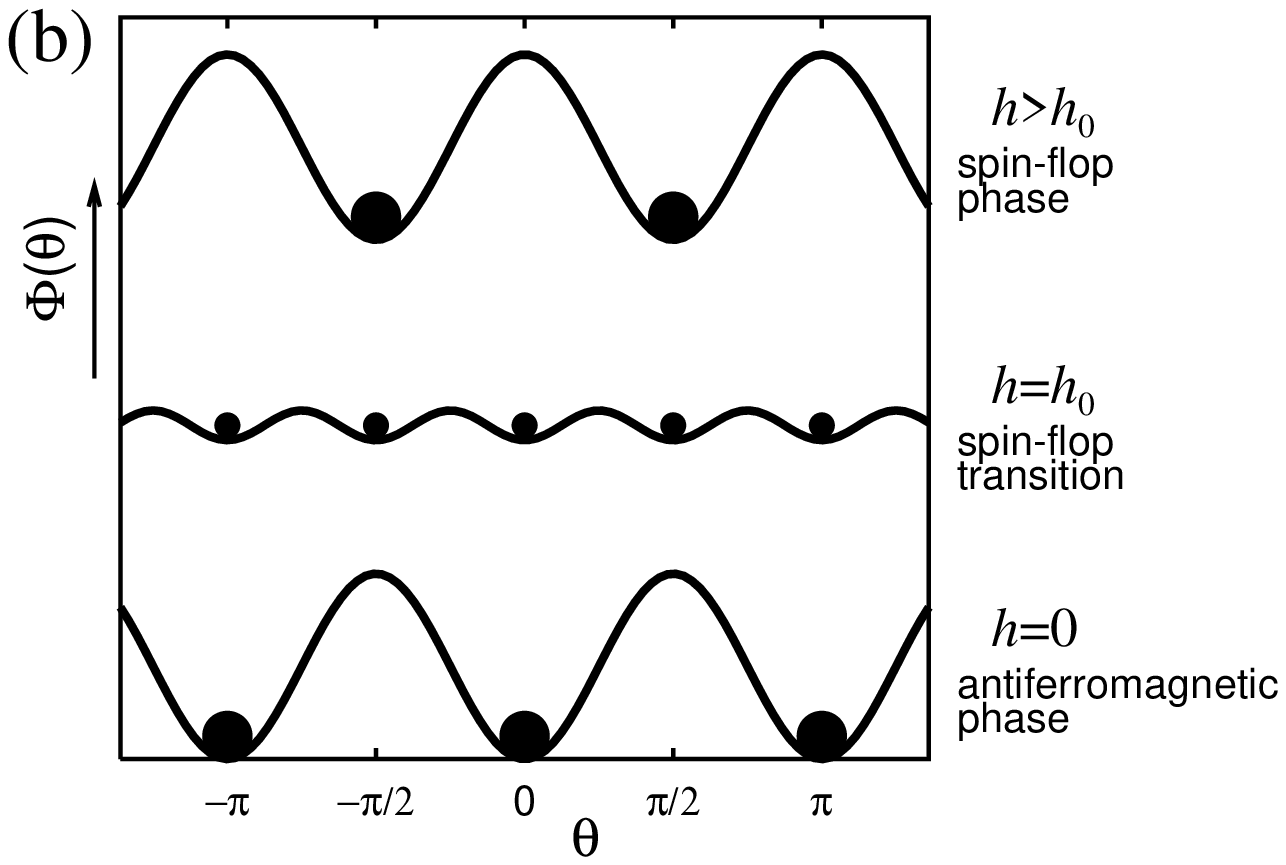}
\caption{
\label{fig3ab}
Schematic evolution of the potential 
profiles ${\Phi(\theta)}$ 
under influence of the applied field for easy-axis systems ($K>0$)
in an oblique field (a) and 
in a field in direction of the tetragonal axis (b).\\
In (a) for the canted phase two different barriers 
occur between equivalent equilibrium states.
Different domain walls correspond to these barriers: DW~I 
between state (1) and (2), and DW~II between  (1) and (3).
}
\end{figure}
The relative orientation of the vectors ${\mathbf l}$ 
and ${\mathbf m}$ is fixed by the sign of the constant $d$. 
This handedness of
the magnetic structures reflects the chiral character 
of the interaction (\ref{Dz}) and leads 
to the nonequivalence of energies for 
the states with antiparallel values of ${\mathbf l}$ 
in oblique magnetic fields.
One can understand this considering the shapes
of the potential ${\Phi}$ (\ref{density00}) 
with applied fields (Fig.~\ref{fig3ab}).
In the general case (Fig.~\ref{fig3ab}(a)) of an oblique field,
i.e. an applied field deviating from the tetragonal axis,
the antiferromagnetic phase with ${\theta}={\pi}n$
transforms into the WF phase ($h>h_{c}$)
via the canted phase.
From the potential profile for this 
canted phase (Fig.~\ref{fig3ab}(a), $0<h<h_c$)
one immediately sees for a given state
that the corresponding state 
with antiparallel orientation of ${\mathbf l}$ 
has a different energy and generally is 
not an equilibrium state.
Thus, the symmetry between states with
antiparallel ${\mathbf l}$ 
peculiar to ordinary antiferromagnetic phases is violated.
The stable states in this canted phase
are separated by two types of potential barriers. 
%
% Correspondingly 180-degree walls of the antiferromagnetic phase 
% evolve into two different types of domain walls
% (DW~I and DW~II). 
%
At the transition into the WF phase the lower potential barrier 
disappears and the higher separates states
with ${\theta}={\pi}/2+2n{\pi}$.
In a magnetic field along the tetragonal axis (Fig.~\ref{fig3ab}(b))
the potential barriers in ${\Phi}$ (\ref{density00}) 
disappear as the field approaches the spin-flop field
from both sides, i.e. in the antiferromagnetic (${h<h_0}$) 
and in the spin-flop phases (${h>h_0}$).
This means that at the transition field ${h=h_0}$ 
the potential barriers between coexisting antiferromagnetic 
and spin-flop states are anomalously low and are determined 
by the values of the fourth-order anisotropy.

\subsubsection{$K<0 \qquad$      Easy-plane system} \label{easyplanesys}
In the ground state  for easy-plane systems $K<0$
the staggered magnetization lies in the basal plane
with a spontaneous magnetization, ${|{\mathbf m}|=d/{\lambda}}$, 
perpendicular to the vector ${\mathbf n}$.
Therefore, this is a {\textit {weak-ferromagnetic phase}}.
The behaviour of the system 
under influence of 
a magnetic field is similar
to that in the easy-axis  system (section \ref{easyaxissys})
for $h_z $ larger than the spin-flop field, 
namely the vector ${\mathbf l}$ is oriented
perpendicular to ${\mathbf h}$, and the total magnetization gradually
increases in increasing field.
\subsection{Helical structures}\label{Helical structures}
The equations minimizing the functional  
(\ref{energyT}) also permit solutions 
with chiral modulations
propagating in the basal plane. 
First we consider structures
modulated along a certain fixed direction in the basal plane
and homogeneous perpendicular to this direction. 
This yields one-dimensional spirally modulated 
states comprising {\textit {helicoids}} and {\textit {cycloids}}. 
In the absence of in-plane anisotropy
all propagation directions are equivalent. 
\begin{figure}
\includegraphics[width=8.0cm]{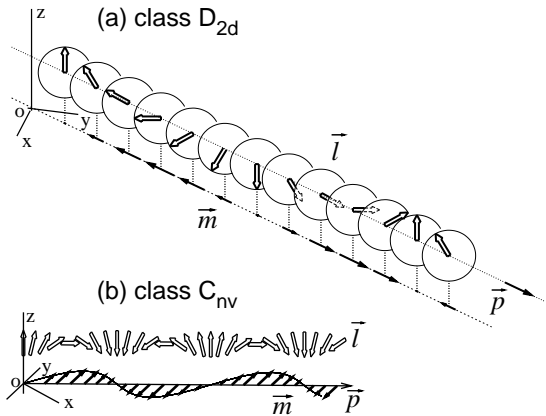}
\caption{
\label{fig4ab}
Basic modulated structures 
(a) - helicoid for systems with $D_{2d}$ symmetry; 
(b) - cycloids for antiferromagnets with $C_{nv}$ symmetry.
}
\end{figure}
The structure of these modulated states depends 
on the crystal symmetry which manifests itself
in different functional forms of the Lifshitz invariants
(see (\ref{lifshitzC}), (\ref{lifshitzD})).
For antiferromagnets belonging to the crystallographic class $D_{2d}$ 
the staggered vector ${\mathbf l}$ rotates 
in the plane perpendicular
to the propagation direction, i.e. ${\psi}={\pi}/2$.
These states are  {\textit {helicoids}} (Fig.~\ref{fig4ab}(a)). 
This rotation of ${\mathbf l}$ reminds the behaviour of 
the magnetization vector for Bloch walls 
in ferromagnets (see \cite{Hubert98}).
In the case of $C_{nv}$ symmetry,
${\mathbf l}$ rotates in the plane formed by the tetragonal axis 
and the propagation direction (${\psi}=0$) 
forming {\textit {cycloids}} (Fig.~\ref{fig4ab}(b)).
This is akin to N{\'e}el domain walls in ferromagnets. 
The rotation in the spirals has a fixed sense determined 
by the condition that the inhomogeneous Dzyaloshinskii-Moriya 
energy must be negative.
In both cases, the spirals are accompanied 
by oscillations of 
the magnetization ${\mathbf m}$ 
perpendicular to the plane of rotation according to 
(\ref{m}) (Fig.~\ref{fig4ab}).
In zero field, $m=d\sin{\theta}/{\lambda}$.\\
Under the influence of an applied magnetic field 
the spirals orientate in such a way
that the rotation of ${\mathbf l}$ 
occurs 
in the plane perpendicular 
to the projection of ${\mathbf h}$ onto the basal plane.
As above for the homogeneous structures, 
we assume that the magnetic
field lies in the $XOZ$-plane and the vector 
${\mathbf n}$ in the $YOZ$-plane.
For {\textit {helicoids}} ($D_{2d}$ symmetry) the propagation direction 
is along the $x$--axis and for {\textit {cycloids}} ($C_{nv}$ symmetry) 
along the $y$--axis.
The spatial coordinate along
the propagation direction is measured 
in reduced units of ${x_0}$ 
according to (\ref{units}),
${\xi}=x_i/x_0$.
In these reduced units the energy functional (\ref{energyT}) for
one-dimensional modulations assumes the form
\begin{equation}
\widetilde{\widetilde{W}}=%
\sqrt{\frac{A|K|}{\lambda}}\int\Bigg\{\left(\frac{d{\theta}}
{d {\xi}}\right)^2 +{\Phi}({\theta})+\frac{4D}{{\pi}D_0}
\left(\frac{d{\theta}}{d{\xi}}\right)\Bigg\}\,d{\xi}\,,
\label{energyS}
\end{equation}
where the multiplicative constants 
due to the integrations in directions of $z$
and perpendicular to $\xi$ are absorbed in $\widetilde{\widetilde{W}}$.
${\Phi}({\theta})$ is given by (\ref{density00}) 
and $D_0$ by (\ref{units}).
The first integral of the Euler equation for the functional 
(\ref{energyS}) is readily derived,
\begin{equation}
\left(\frac{d{\theta}}
{d{\xi}}\right)^2 -{\Phi}({\theta}) = E\,,
\label{First}
\end{equation}
where $E$ is an integration constant. 
In passing, we remark that the Euler equation 
with the potential 
${\Phi}({\theta})$ from (\ref{density00})
is related to the {\textit {double sine-Gordon}} equation.
\cite{DSG-eq}

\begin{figure}
\includegraphics[width=8.0cm]{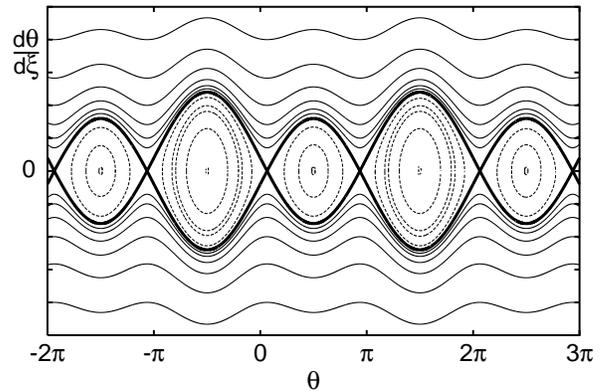}
\caption{
\label{fig5}
Typical phase portrait 
of the solutions for the equation (\ref{First}) 
in the canted phase. The separatrix curves between
open (continuous lines) and closed (dashed lines) orbits 
are highlighted by a thick line.
% with ${\nu}=0.3$.
}
\end{figure}
\begin{figure}
\includegraphics[width=8.0cm]{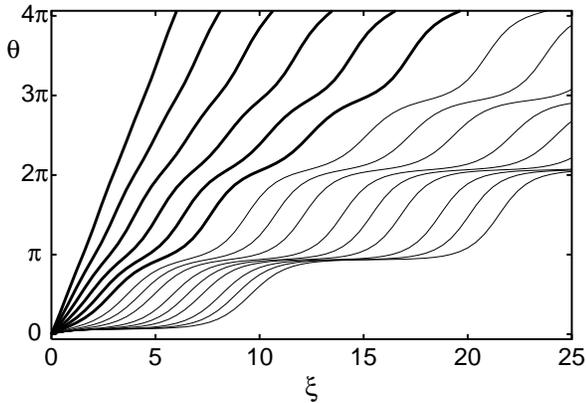}
\caption{
\label{fig6}
Series of solutions of (\ref{First})
${{\theta}}({\xi}, E)$ 
corresponding to the open orbits in Fig.\ 5. 
Solutions shown by thinner lines correspond 
to open orbits with $E>0$ close 
to the separatrix curve of Fig.\ 5 (not shown there). 
}
\end{figure}
Typical phase trajectories of (\ref{First}) 
in the (${\theta}_{\xi}$, ${\theta}$)-phase plane
are plotted in Fig.~\ref{fig5}
(here, we use the abbreviation 
${\theta}_{\xi} \equiv d{\theta}/d{\xi}$ 
). 
The separatrices obtained for $E=0$
cross each other in the points corresponding to the minima
of the function ${\Phi}({\theta})$.
They divide the phase plane 
into regions with {\textit {closed}}
($E<0$) and {\textit {open}} ($E>0$) trajectories (Fig.~\ref{fig5}).
The closed trajectories correspond to alternating rotation 
of the staggered vector and obviously are not of interest 
for our model as they describe inhomogeneous states 
with alternating sense of rotation 
that do not minimize the energy related 
to the inhomogeneous Dzyaloshinskii-Moriya interactions.
The modulated states with fixed rotation sense 
are described by {\textit {open}} trajectories. 
The integration of eq.~(\ref{First}) 
with ${\Phi}({\theta})$ from (\ref{density00}) 
yields the set of solutions ${{\theta}}({\xi}, E)$ 
parametrized by the constant $E$ (Fig.~\ref{fig6}).
These solutions can be expressed analytically 
as certain cumbersome combinations of elliptic 
functions \cite{Korn} 
(see, e.g., Refs.~\onlinecite{Dz64,AFM89,FNT99}).
Here, for simplicity, we derive representative solutions 
by direct numerical integration of (\ref{First}).
Using (\ref{First}) 
the energy density $\bar{w}$ 
averaged over a period ${\Xi}$ of a modulated state 
can be written as
functions of the parameter $E$
\begin{eqnarray}
\bar{w} & = & \frac{1}{{\Xi}(E)}\int_0^{2{\pi}}%
\frac{[E+2{\Phi}({\theta})]
d{\theta}}{\sqrt{{\Phi}%
({\theta})+E}}-\frac{2{\pi}D}{{\Xi}(E)},
\nonumber\\
& {\Xi}(E) & = \int_0^{2{\pi}}%
\frac{d{\theta}}{\sqrt{{\Phi}({\theta})+E}}\,.
\label{SpiralE}
\end{eqnarray}
Note, eq.~(\ref{First}) does not involve contributions
from the inhomogeneous chiral Dzyaloshinskii-Moriya interactions
and, thus, its solutions (Fig.~\ref{fig6})
do not depend on these chiral interactions.
Therefore, these solutions (Fig.~\ref{fig6}) 
have the same functional form 
as those for the corresponding model 
with $D=0$ (centrosymmetric systems). 
However, the energy $\bar{w}$ (\ref{SpiralE}) of the system
depends on the contribution from the Lifshitz invariants. 
This energy has different values 
for different integral curves
${{\theta}}({\xi}, E)$.
The equation $d\bar{w}/(dE)=0$ 
to derive the optimal values $\tilde{E}$ 
can be reduced to the following form
\begin{equation}
\int_0^{2{\pi}}d{\theta}\sqrt{{\Phi}({\theta})+E}=4D/D_0\,.
\label{EqC}
\end{equation}
Hence, the spiral structure
described by the integral curve 
${{\theta}}({\xi},\tilde{E})$ obeying (\ref{EqC})
corresponds to the equilibrium magnetic structure
realized in a noncentrosymmetric system,
where inhomogeneous Dzyaloshinskii 
interactions $w_D$ are operational.
From the solutions of eqs.~(\ref{First}) and (\ref{EqC}) 
the other equilibrium parameters of the spiral structures 
are readily calculated. 
In particular,
eq.~(\ref{SpiralE}) yields the period of 
the structure ${\Xi}$,
and the oscillating components of the vector ${\mathbf m}$ 
are expressed via (\ref{m}) 
as functions of ${{\theta}}({\xi}, \tilde{E})$.
Depending on the ratio
${D/D_0}$ for the relative strength of the chiral interactions
the modulated structures display 
the following characteristic evolution:
For strong chiral interactions
${D/D_0} \gg 1$ 
the influence of the energy (\ref{density00}) 
is negligible and 
in the equilibrium states the staggered vector
rotates with an essentially 
fixed ``velocity'' ${\theta}_{\xi}$
corresponding to the phase trajectories with 
${{\theta}_{\xi} =2D/({\pi}D_0)}$.
% It is important to mention 
% that 
Perturbations of the uniform 
rotation for the spirals are related to the shape of the 
potential profiles 
(cf. Fig.~\ref{fig4ab} and the open trajectories in Fig.~\ref{fig5}).
Thus, the functional dependencies of 
${{\theta}}({\xi}, \tilde{E})$ 
contain important information
on internal interactions of a system. 
When values of the parameter are smaller, ${{D/D_0}\simeq 1}$  
the influence
of the potential ${\Phi}({\theta})$, 
which determines preferable orientations in the crystal, 
violates uniform rotation of  the staggered magnetization
${\mathbf l}$ in a spiral. 
Further weakening of the chiral
interactions leads to ``pinning'' of ${\mathbf l}$
along certain (``easy'') directions and squeezes the regions 
with ``disadvantageous'' orientations of the vector ${\mathbf l}$. 
This tendency results in the formation of structures
consisting of large domains with homogeneous states
separated by  narrow transition regions in which the vector ${\mathbf l}$
rotates from one easy direction to another similar to domain walls. 
Finally, for a certain critical value of $D$ 
the modulated phase is transformed into the homogeneous state.
This transition is signalled by  an unlimited 
growth of the period for the modulated state.
In the phase space, states at 
this transition into the homogeneous state 
correspond to the separatrix (Fig.~\ref{fig5})
which describes a set of isolated domain walls, 
i.e. walls with infinite separation between them. 
The finite stiffness of the exchange 
interaction prevents a complete annihilation of 
the domain walls and they may exist with finite thickness 
within homogeneous states as metastable topologically 
stable objects acting as nucleation 
centers during a reversal transition 
from the homogeneous into the modulated state.

Finally, it should be stressed 
that eqs.~(\ref{First}) and (\ref{EqC}) 
provide general and rigorous solutions 
for one--dimensional modulated structures 
in magnets with Lifshitz invariants
of type (\ref{lifshitz}) with arbitrary functional 
form for the potential  
${\Phi}({\theta})$ in the functional (\ref{energyS}). 
The above described evolution of the modulated states 
is not restricted 
to any particular form of ${\Phi}({\theta})$;
the qualitative picture of this evolution rather has universal 
character for physically reasonable choices for 
${\Phi}({\theta})$.
Particular cases for such chiral spirals 
have been investigated starting from 
the paper \cite{Dz64} for several groups 
of helical ferro- and antiferromagnets.
\cite{Izyumov84,JETP89,AFM89,Nikos01}

\subsection{Stability limits of the modulated states}\label{Stability}

At the transition into homogeneous states 
the chiral spirals disintegrate into a system
of noninteracting planar domain walls. 
Such a transition can be found by
comparing the energy of the chiral spiral to
the energy of domain walls separating 
regions with different homogeneous states of the system.
\cite{JETP89,AFM89}
Below in section \ref{Localized states}, we will 
discuss domain walls as topological defects 
in the magnetic system which require an excitation energy.
Here, we are concerned with the competition between 
homogeneous and modulated equilibrium states. 
Then, a gain of energy through proliferation of domain walls 
indicates the instability of homogeneous states compared
to a modulated state.

Let us consider a planar isolated domain wall 
between two infinitely extended regions 
with different spatially homogeneous magnetic structures 
that are described by the functional (\ref{density00}),
i.e. equilibrium states in Fig. \ref{fig3ab}.
The equilibrium structure of 
this isolated wall is determined
by solving (\ref{First}) with the boundary conditions 
${\theta}\!\mid_{{\xi}=\pm\infty}={\theta}_{1,2}$,
$(d{\theta}/d{\xi})_{{\xi}=\pm\infty}=0$,
where ${\theta}_{1,2}$ are homogeneous configurations
determined by 
${{\Phi}_{\infty} \equiv \min [{\Phi}({\theta})]=
{\Phi}({\theta}_{1}) = {\Phi}({\theta}_{2})}$
for the functional (\ref{density00}).
The direct integration
of (\ref{First}) yields the following results for
the dependence of ${\theta}({\xi})$ 
in the wall and for the domain wall energy ${\sigma}$ 
(see \cite{Bar88,Hubert74})
\begin{equation}
{\xi}-{\xi}({\theta}_1)= \int_{{\theta}_1}^{{\theta}}
\frac{d{\theta}'}{\sqrt{[{\Phi}({\theta}')-%
{\Phi}_{\infty}]}}\,,
\label{wallSt}
\end{equation}
\begin{equation}
{\sigma}= \frac{{\pi}}{2}D_0\int_{{\theta}_{1}}^{{\theta}_{2}}
\sqrt{[{\Phi}({\theta})-{\Phi}_{\infty}]} \,d{\theta}\,
\pm D|{\theta}_{1}-{\theta}_{2}|\,.
\label{wallEnergy1}
\end{equation}
The function ${[{\Phi}({\theta})-{\Phi}_{\infty}]}$ 
is the deviation of the energy density (\ref{density0}) 
from the minimal value ${\tilde{w}_{\infty}}$ corresponding 
to the homogeneous states in adjacent domains.
The first term in (\ref{wallEnergy1}) is positive
and represents increased energy contributions 
compared to those of the homogeneous states 
${\theta}_1$, ${\theta}_2$.
This increase is due to inhomogeneous exchange interactions 
and interactions included 
into the functional ${\Phi}$ (\ref{density00}). 
These defect energies are typical for 
the energy of magnetic domain walls.\cite{Hubert98}
The second term is specific 
for {\textit {noncentrosymmetric}} systems.
Its sign is determined by the rotation sense of 
the staggered vector ${\mathbf n}$ in the domain wall.
Clearly, for any sign of the constant $D$ there 
exists a rotation sense of the staggered magnetization 
leading to {\textit {negative}} values for this energy contribution 
and, consequently, to a decrease of the domain wall energy. 
For sufficiently 
strong inhomogeneous chiral Dzyaloshinskii-Moriya interactions 
the total energy of the domain wall may be negative compared 
to the energy of homogeneous states.
This manifests an instability 
of the homogeneous state 
with respect to chiral modulations. 
As already discussed above, such a transition
takes place for coefficients $D$ larger than a certain 
threshold value necessary to overcome 
the positive energy contribution 
for inhomogeneous states 
due to the conventional magnetic interactions.

\begin{figure}
\includegraphics[width=8.0cm]{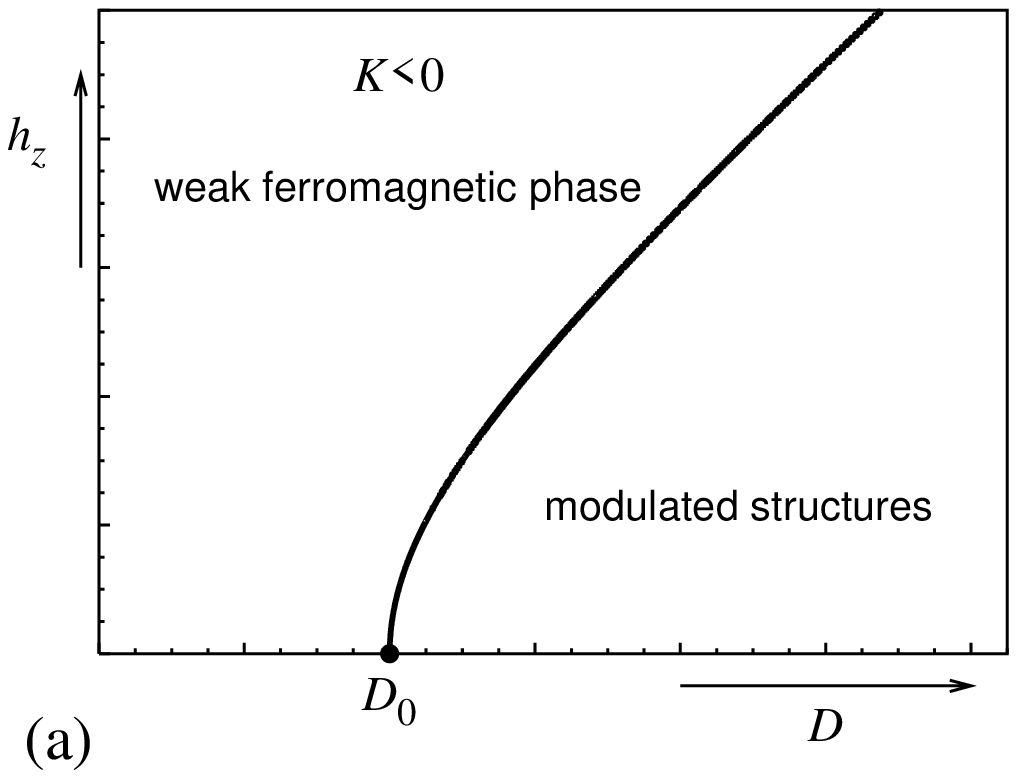}
\includegraphics[width=8.0cm]{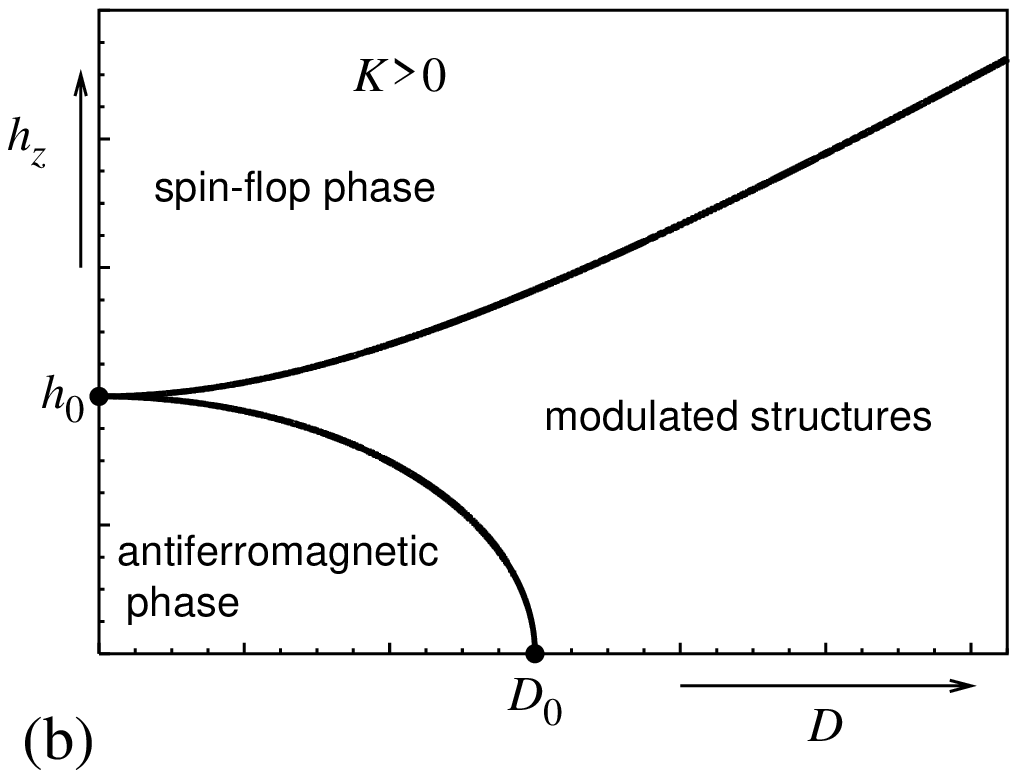}
\caption{
\label{fig7ab}
Magnetic phases 
in the magnetic field along the
tetragonal axis in 
dependence on strength $D$ of the inhomogeneous
Dzyaloshinskii-Moriya interactions (a) $K<0$; (b) $K>0$. 
Note that for easy-axis antiferromagnets (b) 
in the vicinity of spin-flop field the modulated phases 
exists at arbitrarily 
small values of $D$.
}
\end{figure}
 
The wall energy (\ref{wallEnergy1}) 
can be expressed via
the height of the potential barrier,
\({{\Delta}{\Phi}=%
\max[{\Phi}({\theta})]-{\Phi}_{\infty}}\),
that separates the equilibrium states
${\theta}_{1}$ and ${\theta}_{2}$.
This can be written in the following form
\begin{equation}
{\sigma}= \left[{\gamma} D_0
\sqrt{{\Delta}{\Phi}}\pm D\right]
{|{\theta}_{1}-{\theta}_{2}|}\,,
\label{wallEnergy2}
\end{equation}
where ${\gamma}$ is 
a numerical factor 
determined by the average value of 
the integrand in (\ref{wallEnergy1}).
These results have clear physical meaning. 
The higher the energy barrier ${\Delta}{\Phi}$
the stronger the chiral interaction necessary
to overcome it and to stabilize modulated states.
As was shown above, the potential profile
${\Phi}({\theta})$ (\ref{density00})
strongly depends
on the strength and direction of the applied field (Fig.~\ref{fig3ab}). 
Correspondingly, the critical values $D_c$
of the Dzyaloshinskii-Moriya constant $D$
for transitions between homogeneous ($D<D_c$) 
and modulated chiral states ($D>D_c$) 
vary strongly with an applied magnetic field. 
The critical surfaces for these transitions 
in parameter space are given by the equation ${\sigma}=0$ 
using the functional (\ref{density00}).
For easy-plane systems ($K<0$) the critical surface is
\begin{equation}
\frac{D_c}{D_0}= \sqrt{\left|\frac{h_z^2}{K}-1\right|}
\left[\sqrt{1-{\nu}}-\frac{{\nu}}{2}\ln\left(
\frac{\sqrt{1-{\nu}}+1}{\sqrt{1-{\nu}}-1}\right)\right]\,.
\label{boundary3}
\end{equation}
This equation also describes 
the boundaries of the modulated states
for easy-axis systems ($K>0$) 
in magnetic fields larger than the spin-flop field ($h_z > h_0$ ).
For lower fields ($h_z < h_0$ ) the equation
 \begin{equation}
\frac{D_c}{D_0}= \sqrt{1-\frac{h_z^2}{h_0^2}}
\left[\sqrt{1-{\nu}^2}+{\nu}\arcsin{\nu} \right],
\label{boundary1}
\end{equation}
gives the transition into the canted
phase (${\nu} \le 1$). Finally, the equation
\begin{equation}
\frac{D_c}{D_0}= \sqrt{1-\frac{h_z^2}{h_0^2}}
\left[\sqrt{{\nu}-1}+{\nu}\arcsin\sqrt{\frac{1}{{\nu}}} \right],
\label{boundary2}
\end{equation}
describes the transition into 
the weak-ferromagnetic phase (${\nu} > 1$).
In particular, at zero field 
the critical value $D_c$ equals $D_0$. 
Thus, this constant is the lowest value
for the Dzyaloshinskii constant $D$ 
to induce modulated ground states.
The ($D$,$h_z$) phase diagrams 
for an applied field in direction of the tetragonal
axis are shown in Fig.~\ref{fig7ab} for 
the two cases 
$K\,\stackrel{\textstyle{<}}{>}0\,$.
 
\begin{figure}
\includegraphics[width=8.0cm]{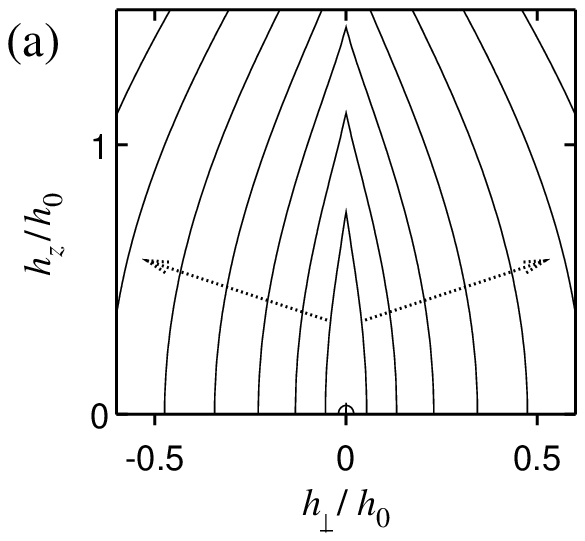}
\includegraphics[width=8.0cm]{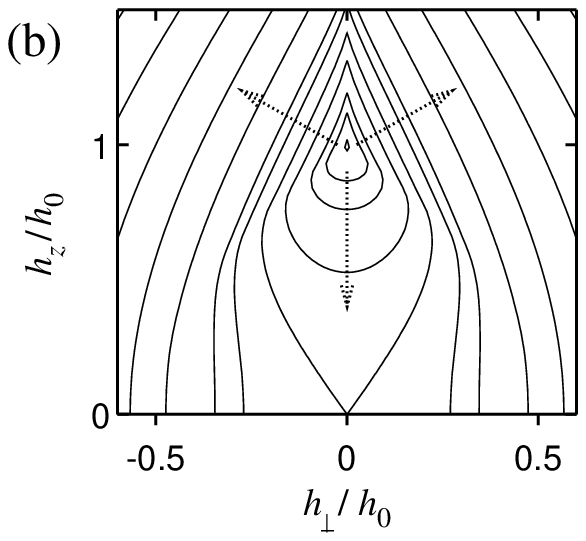}
\caption{
\label{fig8ab}
Contour lines $D_c$($h_{\perp}$, $h_z$)=const 
of the critical surfaces for modulated states.
(a) $K<0$ surface given by (\ref{boundary3}); 
(b) $K>0$ surface according 
to (\ref{boundary3}), (\ref{boundary1}), 
and (\ref{boundary2}). 
(The direction of increasing $D_c$ is indicated
by dotted arrows.)
}
\end{figure}
For easy-plane systems ($K<0$) 
the critical surface $D_c(h_{\perp}, h_z)$ has a minimum
in the origin with $D_c(0,0)=D_0$
and monotonically increases
with increasing magnetic field for any direction (Fig.~\ref{fig8ab}(a)).
When $D<D_0$ the chiral interactions are too weak to overcome
the pinning due to the uniaxial easy-plane anisotropy.
Then, the system exists in the homogeneous state 
with the staggered magnetization in the basal plane and 
a weak spontaneous magnetization (WF phase).
For $D>D_0$ the WF phase becomes unstable. 
Under the influence of the inhomogeneous chiral interactions 
the vector ${\mathbf n}$ ``escapes'' from
the basal plane and a chiral helix is formed.
We add a remark about 
the peculiarity of this type of helix.
In known easy-plane systems
with  helical structures the magnetization 
(or the staggered magnetization 
in the case of antiferromagnets)
rotates in the ``easy-plane'' and the propagation vector 
is perpendicular to this plane. 
In these noncentrosymmetric magnets 
such spirals are stabilized by Lifshitz 
invariants with gradients along the ``hard-axis''.
In our model, however, the Lifshitz invariants 
include only gradients
in the basal plane. 
Correspondingly, 
the chiral modulations 
in these systems have propagation directions
only in the basal plane. \cite{Izyumov84,JETP89}

For easy-axis systems ($K>0$) 
the critical  surface $D_c(h_{\perp}, h_z)$
has a more involved shape. 
In this case the lowest
value of $D_c$ equals zero. 
This is reached at the spin-flop 
field ${(0,\pm h_0)}$ (Figs.~\ref{fig7ab}(b), \ref{fig8ab}(b)). 
Thus, near the spin-flop transition 
the modulated states
arise at arbitrarily small values of $D$. 
This unusual situation 
is due to the particular evolution of 
the potential profile 
({\ref{density00}) 
in a magnetic field directed 
along the tetragonal axis (Fig.~\ref{fig3ab}(b)). 
In this case the uniaxial anisotropy 
and the applied magnetic field have competing influence
on the magnetic structure.
While the easy-axis anisotropy orientates 
the staggered magnetization
along the tetragonal axis,
the applied field orientates it
perpendicular to this axis. 
An increasing magnetic field 
in the region $h<h_0$ 
gradually decreases the potential barrier between 
the states of the antiferromagnetic phase 
with antiparallel staggered magnetization
(Fig.~\ref{fig3ab}(b)). 
When the spin-flop is reached 
the applied field completely cancels
the influence of uniaxial anisotropy,
and the potential (\ref{density00}) 
equals zero for any orientation 
of the staggered magnetization. 
This infinite degeneracy of magnetic 
states is artificial because of 
the neglect of higher order anisotropy contributions. 
A fourth--order 
uniaxial anisotropy $K_2\sin^4{\theta}$ 
removes this degeneracy.
We may generally state 
that in noncentrosymmetric easy-axis antiferromagnets 
near the spin-flop field the potential barrier 
between minima (and, therefore, the critical values of $D$) 
is determined by the much weaker fourth-order anisotropy 
constant ${K_2}$. 
We add two remarks here:
(i) For centrosymmetric easy-axis antiferromagnets
(models I.) and II.) of subsection \ref{Energy}),
the situation near spin-flop transitions
is physically comparable and
has been studied in detail.\cite{Bar88,FNT86}
There, the relation between interactions 
and homogeneous magnetic states is well understood.
(ii) In cubic helimagnets,\cite{Lebech89}
where magnetocrystalline anisotropy 
is represented only by fourth-order terms, 
no suppression of the modulated states has been observed,
rather chiral modulations exist 
in the complete region of existence 
of magnetically ordered states. 
Estimates based on the physical origin of 
these magnetic energy contributions yield
so weak threshold values 
that modulated chiral states 
near the spin-flop field should be expected generally.\cite{AFM89}
Therefore, this competition 
between antisymmetric exchange and anisotropies 
makes the easy-axis noncentrosymmetric 
antiferromagnets particularly interesting systems
for a search for 
and investigation of modulated chiral states.

\section{Localized chiral structures}\label{Localized states}
In this section we consider 
the influence of the Dzyaloshinskii-Moriya 
interactions on localized magnetic defects within
homogeneous magnetic configurations. 
As there is a wide
variety of possible defect structures 
in the different phases,
we will present only a few examples 
to demonstrate the general principles 
which rule defect structures for the noncentrosymmetric 
antiferromagnets under the influence of 
chiral couplings.
%\cite{unpublished results by Bogdanov 02}
A formal mathematical description of isolated,
planar one-dimensional defect structures, 
i.e. domain walls, was already developed above 
in section \ref{Stability}. 
In the next subsection, 
we discuss physical importance and properties of such domain walls 
for antiferromagnetic systems described by our model.
Subsection \ref{vortices} is devoted to
linear two-dimensional defect structures, i.e. vortices.

\subsection{Domain walls or kinks}\label{DomainWalls}
Planar defects (domain walls or {\textit {kinks}}) are commonly
observed magnetic localized states in many classes of 
antiferromagnetic materials.\cite{AFdomains}
They separate homogeneous states 
with different degenerate 
directions of the staggered magnetization.
An example is provided by 180-degree domain walls
between regions with antiparallel staggered magnetization,
i.e. different antiferromagnetic phases,  in easy-axis antiferromagnets. 
In the antiferromagnets under discussion, the rotation of the vector
${\mathbf n}$ within a domain wall is accompanied by oscillation of 
the total magnetization. 
The spin arrangement in such domain walls is similar to
that in the corresponding spirals (Fig.~\ref{fig4ab}). 
Rotation of ${\mathbf n}$ as in a Bloch wall
with longitudinal modulation of the vector ${\mathbf m}$ (Fig.~\ref{fig4ab}(a)) 
should occur in noncentrosymmetric antiferromagnets 
belonging to crystallographic class $D_{2d}$.
N{\'e}el-wall-like structures 
with transversal oscillation of the magnetization 
correspond to the noncentrosymmetric antiferromagnets
from class $C_{4v}$ (Fig.~\ref{fig4ab}(b)).
The  inhomogeneous chiral Dzyaloshinskii-Moriya interactions
do not influence the structure of the domain walls
(see above, remarks following eq.(\ref{SpiralE})).
% in section~\ref{Helical structures}).
However, the domain
wall energies do depend 
on the rotation sense according to (\ref{wallEnergy2}). 
As discussed above, 
the modulated structures have a fixed sense of rotation
that corresponds to a decrease of total energy 
compared to the homogeneous states.
Spirals with opposite sense of rotation 
are unstable (even with respect to
the homogeneous states) and never arise in real systems. 
Contrary to this, domain walls 
with disadvantageous sense of rotation, 
although increasing the energy, 
should be found within these antiferromagnets 
with similar probability
because domain walls in antiferromagnets 
have mostly ``kinetic'' origin
in contrast to domain structures in ferromagnets,
i.e. antiferromagnetic domain structures 
are formed during the transition to the ordered states 
or as a result of reorientation transitions.
These processes are largely independent 
of domain wall energies.

\begin{figure}
\includegraphics[width=8.0cm]{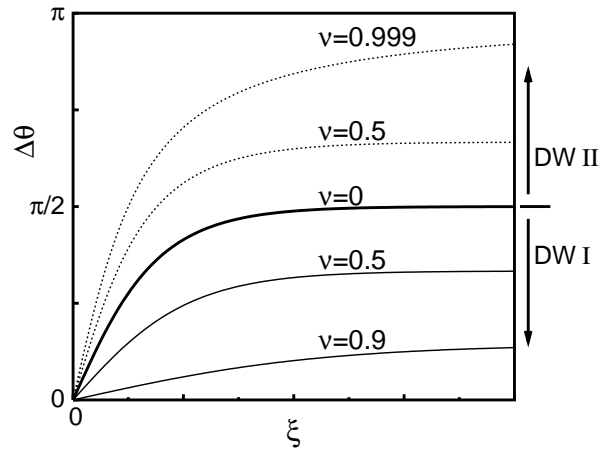}
\caption{
\label{fig9}
Structure of domain walls in the canted phase 
given by the turn angle ${\Delta}{\theta}$ 
as function of distance from the wall-center $\xi=0$
for different values of the parameter ${\nu}$. 
The 180-degree
domain walls of the AF phase (${\nu}=0$) 
are deformed in the canted
phases either into walls with a decreased value 
of ${\Delta}{\theta}(\xi \rightarrow \infty)$
(DW~I, profiles with continuous lines) or 
an increased ${\Delta}{\theta}(\xi \rightarrow \infty)$
(DWII, profiles with dashed lines). 
DW~II are transformed into  360-degree walls 
in the limit ${\nu}=1$.
}
\end{figure}
\begin{figure}
\includegraphics[width=8.0cm]{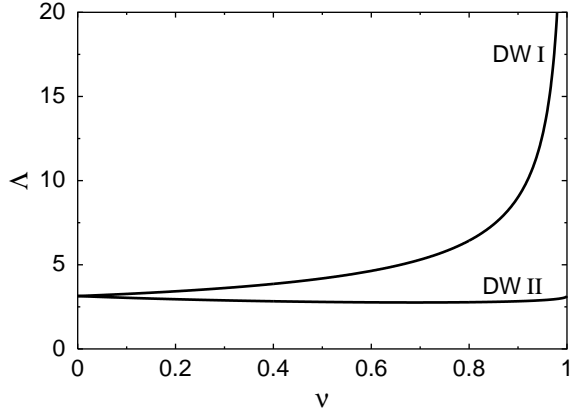}
\caption{
\label{fig10}
The domain wall widths ${\Lambda}$ 
as functions of the parameter ${\nu}$ for the two wall types in 
the canted phase, (a) DW~I (b) DW~II, according to (\ref{wallWidth}).
}
\end{figure}
The structure of such domain walls 
can be derived by integration
in (\ref{wallSt}). 
Here, we restrict ourselves to one example
of a practical calculation. We obtain
the structure and characteristic parameters of the domain walls 
in the canted phase (see Figs.~\ref{fig1abcd}, \ref{fig2}). 
Within all regions of their
existence (${\nu} \le 1$) 
the equilibrium states are separated
by two types of barriers in 
the potential profile ${\Phi}(\theta)$ (Fig.~\ref{fig3ab}(a)). 
Correspondingly there exist two types
of domain walls in the canted phase. 
The first low energy
domain wall (DW~I) separates homogeneous states 
with ${\theta}=\arcsin{\nu}$ and
${\theta}={\pi}-\arcsin{\nu}$; 
and DW~II corresponding to 
the higher potential barrier separates 
states with ${\theta}=\arcsin{\nu}$  
and ${\theta}=-{\pi}-\arcsin{\nu}$ (Fig.~\ref{fig3ab}(a)).
Evaluating the integral (\ref{wallSt}) 
with ${\Phi}$ from (\ref{density00}) 
yields the following results
\begin{equation}
\sin{\theta}=%
\frac{{\nu}\cosh({\xi}\sqrt{1-{\nu}^2}) \pm 1}
{\cosh({\xi}\sqrt{1-{\nu}^2}) \pm {\nu}} \,.
\label{cantedSt}
\end{equation}
Wall structures
for both domain wall types in the canted phase 
with varying $\nu$ 
are displayed in Fig.~\ref{fig9}. 
The effective thickness of domain 
walls ${\Lambda}$ is usually determined as
a distance between points where the tangent 
at the inflection point intersects 
the lines ${\theta}= {\theta}_1$ and
${\theta}= {\theta}_2$.\cite{Hubert98}
For our example this definition yields 
the following expression 
\begin{equation}
{\Lambda} = |{\theta}_1 - {\theta}_2|\left(\frac{d{\theta}}
{d{\xi}}\right)_{{\xi}=0}^{-1}=
\frac{{\pi} \mp 2\arcsin{\nu}}{1 \mp {\nu}}\,.
\label{wallWidth}
\end{equation}
(In (\ref{cantedSt}) and (\ref{wallWidth}) 
the upper/lower signs correspond to
DW~I/DW~II.) 
The dependence of the wall thickness ${\Lambda}$ on $\nu$ for 
both types of walls is shown in Fig.~\ref{fig10}. 
For increasing ${\nu}$ the difference
between magnetic configurations 
in the adjacent domains separated by DW~I 
(${\Delta}{\theta}= {\pi}-2\arcsin{\nu}$)
and the potential barrier (${\Delta}{Phi}= {\Phi}({\pi}/2)-
{\Phi}({\theta}_1) \sim (1-{\nu})^2$)
gradually decreases 
while the thickness of the wall increases. 
At the critical point of the transition 
into the weak ferromagnetic phase, ${\nu}=1$, 
the difference between magnetic states in the domains disappears
and the wall spreads out without bounds. 
For DW~II the potential barrier
${\Delta}{\Phi} \sim (1+{\nu})^2$) 
and ${\Delta}{\theta}= {\pi}+2\arcsin{\nu}$ 
increases with increasing ${\nu}$. 
At the critical point, ${\nu}=1$, these walls
transform into 360-degree domain walls (Figs.~\ref{fig9} and \ref{fig10}). 

Structures and parameters for 180-degree domain walls 
in the antiferromagnetic and spin-flop phases 
can be derived in a similar way.
%\cite{unpublished results by Bogdanov 02}
All these domain walls may play the role of nucleation centers 
during the transition from the homogeneous to modulated states.
On the other hand, as demonstrated in the previous section,
at a transition into the homogeneous state 
the spiral states break down into a system of isolated plane walls.

\subsection{Vortices or skyrmions}\label{vortices}
\begin{figure}
\includegraphics[width=8.6cm]{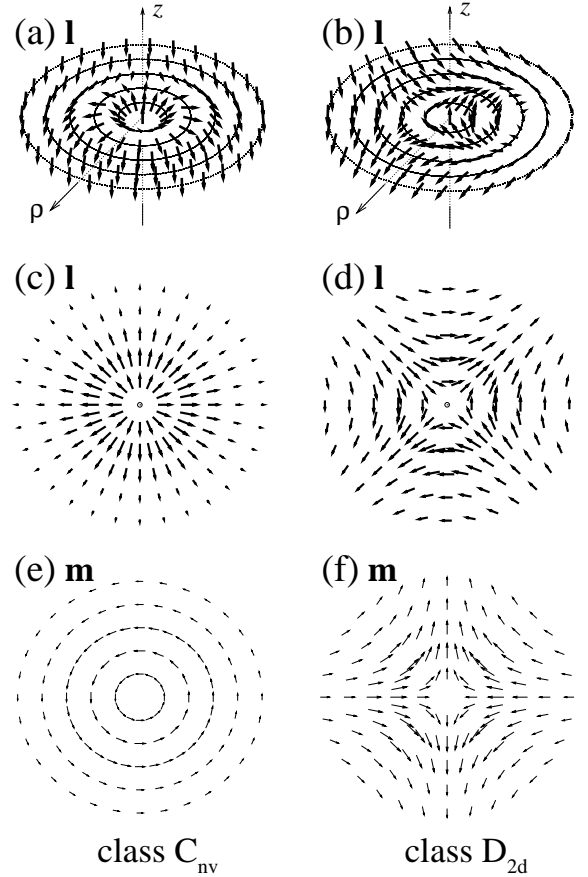}
\caption{
\label{fig11abcdef}
Vortex structure 
for antiferromagnets 
with $C_{nv}$ symmetry (a), (c),(e) 
and  $D_{2d}$ symmetry (b), (d), (f).
(a) and (b) 
distributions of staggered vector ${\bf l}$.
(c) -- (f) projections of ${\bf l}$ 
and oscillating total magnetization 
${\bf m}$ in the basal plane, respectively.
}
\end{figure}
Linear magnetic defects are another type of
topological excitations 
that can exist in noncentrosymmetric
magnetic crystals due to the stabilizing effect 
of the inhomogeneous
Dzyaloshinskii-Moriya interactions.\cite{JETP89}
At zero magnetic field and in fields 
applied along the tetragonal axis 
the model (\ref{energy0}) is invariant 
to rotation about the $z$-axis.
Solutions for the vector ${\mathbf n}({\mathbf r})$
axially symmetric in the basal plane
and uniform along the tetragonal axis, i.e.\ {\textit {vortices}},
obey this symmetry.
As an example of such localized states we consider 
an isolated vortex in the antiferromagnetic
phase ($K>0$, ${\mathbf h}$ is parallel to the tetragonal axis 
and smaller than the spin-flop field $h < h_0$). 
We assume that the staggered magnetization 
is oriented parallel to
the $z$-axis on the vortex axis and rotates 
into the antiparallel orientation with increasing radial
distance from the vortex core.
It is convenient 
to introduce cylindrical coordinates 
for the spatial variables,
${\mathbf r}=x_0({\rho}\cos{\varphi}, {\rho}\sin{\varphi}, z)$,
in the expression 
for the energy (\ref{energyT}).
(As earlier in the case of the spirals, 
we use length units
$x_0=\sqrt{A{\lambda}/{|K|}}$.) 
The analysis of the energy functional (\ref{energyT}) 
shows that the problem 
has solutions ${\theta}({\rho})$ with
\begin{equation}
{\psi}=%
{\varphi} \;{\textrm {for class}}\; {\textit C}_{nv}\,,%
% \;\mbox{\textrm and}
\quad
{\psi} = {\pi}/2-{\varphi} \;{\textrm {for class}}%
\;{\textit D}_{2d}.
\label{vortex1}
\end{equation} 
For $C_{nv}$ symmetry the solution (\ref{vortex1}) 
describes a cycloid-like rotation of the staggered 
magnetization vector ${\mathbf l}$ 
(Figs.~\ref{fig11abcdef}(a) and (c)). 
In the case of $D_{2d}$ symmetry
the vortex has more 
a sophisticated structure (Figs.~\ref{fig11abcdef}(b) and (d)). 
The rotation of the
staggered magnetization in the vortices 
is accompanied by in-plane oscillations
of the total magnetization  
(Figs.~\ref{fig11abcdef}(e),(f)) as
described by eq.(\ref{m}). 

\begin{figure}
\includegraphics[width=8.0cm]{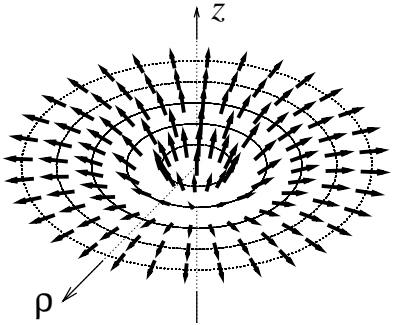}
\caption{
\label{fig12}
Delocalized vortex structure 
in the spin-flop phase in crystals with $C_{nv}$ symmetry.
Staggered vector ${\bf l}$ is shown.
}
\end{figure}
The equilibrium distribution ${\theta}({\rho})$ 
is determined from the differential equation common 
for both classes
\begin{eqnarray}
\frac{d^2{\theta}}{d{\rho}^2} & + & 
\frac{1}{{\rho}}\frac{d{\theta}}{d{\rho}}-
\frac{\sin{\theta}\cos{\theta}}{{\rho}^2}
\label{vortexE}
\\
& & +\frac{4D}{{\pi}D_0}\frac{\sin^2{\theta}}{{\rho}}
-\left(1-\frac{h^2}{h_0^2}\right)\sin{\theta}\cos{\theta}=0
\nonumber
\end{eqnarray}
with boundary conditions 
${\theta}(0) = 0$
and ${\theta}(\infty)=\pi$
for {\textit {localized}} vortices 
in the antiferromagnetic phase 
($h<h_0$, Fig.~\ref{fig11abcdef}),
or ${\theta}(\infty)=\pi/2$ 
for {\textit {delocalized}} vortices 
in the spin-flop phase ($h>h_0$, Fig.~\ref{fig12}).
\begin{figure}
\includegraphics[width=8.0cm]{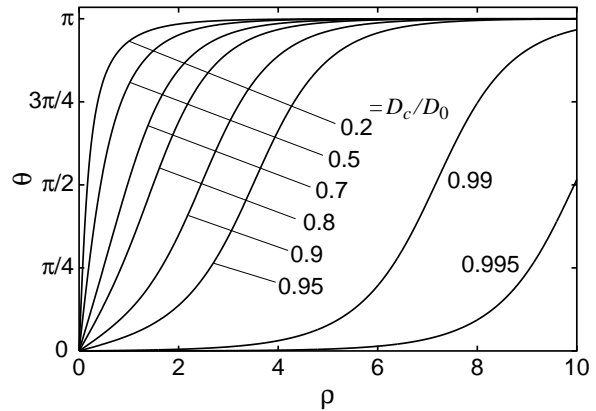}
\caption{
\label{fig13}
Vortex profiles $\theta(\rho)$: solutions of (\ref{vortexE}) 
for different values of $D/D_0$ in zero applied field.
}
\end{figure}
These different boundary conditions 
result in an important physical 
difference between 
these two cases of antiferromagnetic vortices:
For the localized vortices the homogeneous 
equilibrium state is established everywhere
for $\rho\rightarrow\infty$ and the inhomogeneity
is localized in the vortex core.
In the case of the spin-flop phase, 
the vortex structure at $\rho\rightarrow\infty$ 
is inhomogeneous with $\theta=\pi/2$, 
but the angle $\psi$ rotates through 
the a full circle from $0$ to $2\pi$.
Therefore, these vortices are named {\textit {delocalized}}.
Eq. (\ref{vortexE}) has solutions only when $D$ 
is smaller than the critical values for the transition 
to the modulated phase ${D_c(h)}$ 
given by (\ref{boundary1}).
Typical solutions for ${\theta}({\rho})$ 
are plotted in Fig.~\ref{fig13}.
As $D$ approaches the critical value $D_c$ 
the vortex expands without bounds. 
Eq.~(\ref{vortexE}) functionally coincides 
with the equations for isolated vortices 
in other models with Lifshitz invariants,
i.e.\ 
models for noncentrosymmetric ferromagnets, \cite{JETP89,pss94}
for other classes of antiferromagnets,\cite{AFM89}
as well as for chiral liquid crystals.\cite{JETP98}
For detailed analysis of eq.~(\ref{vortexE}) 
and discussion of the related questions see these papers.
 
The vortices or {\textit {skyrmions}}
considered here (Figs.~\ref{fig11abcdef}, \ref{fig12}, \ref{fig13}) 
are non-singular linear defect structures.
They belong to topological defects
studied in many fields of the modern physics.
Similar topological objects arise
in superfluid helium,\cite{Salomaa87} 
in two-dimensional electronic systems (Hall skyrmions),\cite{Sondhi93} 
or in nanomagnetic materials.\cite{PRL01} 
It is important to mention 
that there is a fundamental correspondence 
between these theoretical models.\cite{Pasquier01}
In the isotropic case ($w_D=\tilde{w}=0$ in (\ref{energyT})) 
the equation for the vortex has analytical solutions
which are well-known as  Belavin--Polyakov--solutions 
for nonlinear ${\sigma}$-models.\cite{Belavin75}
These solutions turned out to be unstable 
in centrosymmetric magnetic crystals 
and collapse spontaneously under the influence of 
anisotropic internal interactions or applied magnetic fields.
Thus, the Lifshitz invariants are
crucial for stabilizing these vortex structures
in noncentrosymmetric magnetic crystals.\cite{pss94}
Hence, such low-symmetry magnetic crystals
are interesting and important systems
for investigations of general properties of vortices.

In the spin-flop phase
the vortex states have 
delocalized character (Fig.~\ref{fig12}). 
They are similar to vortex states 
in liquid helium or some textures 
in liquid crystals.\cite{Mermin79 deGennes93}
Such vortices for noncentrosymmetric
antiferromagnets with $d=0$ have been investigated 
in Ref.~\onlinecite{AFM89}.
They readily form localized vortex pairs 
similar to those responsible for Berezinskii-Kosterlitz-Thouless 
transitions.\cite{BKT}
An applied magnetic field deviating from 
the tetragonal axis violates the axial symmetry of the system. 
Then, the two-dimensional localized states are expected 
to have various elongated shapes 
similar to those observed 
in chiral liquid crystals.\cite{Oswald00}
The Lifshitz invariants can also stabilize three-dimensional
localized states (as free spherulites or drops).\cite{pss94}
Up to now no experimental observations or 
theoretical investigations 
of such structures have been reported.

Concluding this section we draw attention 
to an important difference between
the localized states in our model 
and those in other magnetic systems.
In noncentrosymmetric antiferromagnets with weak-ferromagnetism 
due to the oscillating weak magnetization in the basal plane 
the domain walls and the vortices are susceptible 
to the influence of applied magnetic fields. 
For this reason, 
noncentrosymmetric antiferromagnets 
are convenient 
for the study of phase transformations 
and the dynamics of such nonlinear localized excitations.

\section{Relation to experimental observations}\label{Experiment}
The known noncentrosymmetric antiferromagnets 
with weak ferromagnetism
include two tetragonal crystals. 
First, for Ba${_2}$CuGe${_2}$O${_7}$ belonging
to the crystallographic class
$\bar{4}2m$ ($D_{2d}$), 
chiral modulations were discovered five year ago.\cite{Zhelud96}
For the crystallographic class $4mm$ ($C_{4v}$) 
only the antiferromagnet K${_2}$V${_3}$O${_8}$ 
has apparently been investigated and no modulated states 
have been found yet.\cite{Lumsden,Private}
Here, we shortly review experimental data for these 
two antiferromagnetic compounds 
within the framework of our theory. 
In the last subsection we comment on related 
experiments on noncentrosymmetric antiferromagnetic crystals.

\subsection{Ba${_2}$CuGe${_2}$O${_7}$}\label{Ba2CuGe2O7}
A chiral spiral with propagation vector in the basal plane
and period length of about 37 unit cells was found as
magnetic ground state for Ba${_2}$CuGe${_2}$O${_7}$ 
(space group $P\bar{4}2_1m$).\cite{Zhelud96} 
It was also found that 
a rather strong magnetic field applied 
along the tetragonal axis induces a transition 
into a homogeneous state.\cite{Zhelud97}
The field dependencies of the period and
the magnetization reported are 
in quantitative agreement
with theoretical results of Ref.~\onlinecite{Dz64}. 
It appears that there is
no local minimum 
of the period length in dependence on the strength of
a magnetic field applied along the tetragonal axis. 
This implies that the uniaxial anisotropy of this crystal
is of easy-plane type ($K<0$) (cf. Fig.~\ref{fig7ab}). 
To analyze their experimental data the authors explored the
model with $d=0$. 
In their experiments there is no indication 
of effects related to weak ferromagnetism.

Further detailed investigations in magnetic fields
applied along other directions are required to determine the character
of the uniaxial anisotropy and the values of
the other characteristic parameters of the magnetic system 
within the general phenomenological expression (\ref{energy0}) 
for the energy.

\subsection{K${_2}$V${_3}$O${_8}$}\label{K2V3O8}
For this compound, 
at a temperature of 2~K 
(the N{\'e}el temperature is about 4~K) 
the magnetization curves in a magnetic field 
along tetragonal axis and in the basal plane 
indicate reorientation transitions.\cite{Lumsden}
These transitions
are similar to those earlier observed 
in centrosymmetric antiferromagnets
with weak ferromagnetism, 
e.g.\ hematite.\cite{Dejongh74,hematite}
The authors conclude from neutron diffraction experiment
that there are 
no indications of chiral modulations.\cite{Lumsden,Private}
According to the results of our theory
such a situation may take place for easy-axis
systems ($K>0$) with weak chiral interactions ($D<D_0$).
As was discussed above, in the vicinity of the spin-flop field 
the criterion for the stabilization of the modulated states 
is considerably weakened. 
Thus, the search for modulated states in this system
should be started from thorough investigation 
near the spin-flop field.
We add that there are two other 
similar noncentrosymmetric vanadium oxides 
Rb${_2}$V${_3}$O${_8}$ and (NH${_4}$)${_2}$V${_3}$O${_8}$ 
which are supposed to possess antiferromagnetic order 
below 10~K.\cite{Liu95} They could be investigated
in search for effects of chiral interactions.

\subsection{Other noncentrosymmetric antiferromagnets}
\label{OtherSystems-and-CuB2O4}

The copper metaborate CuB${_2}$O${_4}$ (space group 
$I\bar{4}2d$ ($D_{2d}^{12}$))
belongs to the same noncentrosymmetric class 
as Ba${_2}$CuGe${_2}$O${_7}$.
However, according to \cite{Petr99,Roessli,Petr01}
it has a more sophisticated four-sublattice 
antiferromagnetic structure with in-plane anisotropy.
A long-periodic modulated state has been observed in 
this crystal for a certain temperature range.\cite{Roessli,Petr01}
Finally we mention here 
two other noncentrosymmetric antiferromagnets:
a modulated chiral state has been observed 
in the noncentrosymmetric antiferromagnet 
BiFeO${_3}$ (space group $R3c$).\cite{sos}
For CuFeS${_2}$ 
(space group $I{\bar{4}}2d$ ($D_{2d}^{12}$))
antiferromagnetic order was reported to exist, 
however, no details about the magnetic structures
are given.\cite{woolley}

\section{Conclusions}\label{conclusions}
In this paper we show that the novel 
antiferromagnetic crystals Ba${_2}$CuGe${_2}$O${_7}$,
\cite{Zhelud96} and
K$_{2}$V$_{3}$O$_{8}$ \cite{Lumsden}
in spite of the reported difference
in their magnetic properties belong to a common,
previously unknown class of magnetic crystals:
{\textit {noncentrosymmetric antiferromagnets 
with weak ferromagnetism}}.
The phenomenological expression for 
the magnetic energy of such systems
including all interactions allowed by symmetry
(\ref{energy0}) can be reduced to the functional
(\ref{energy2}) which describes the orientation of
the staggered magnetization and can be considered
as general model for two-sublattice antiferromagnets. 
It includes, as specific cases, all main classes
of antiferromagnetic crystals (collinear antiferromagnets, 
antiferromagnets with weak ferromagnetism,
noncentrosymmetric antiferromagnets without and
with weak ferromagnetism). 
Further, by using realistic assumptions 
about the relative strengths of
the phenomenological constants in (\ref{energyT}) 
the problem has been reduced 
to the case that the rotation
of the staggered magnetization 
is restricted to a certain fixed plane. 
This simplification yields a representative 
and realistic approximative model replacing 
the general model (\ref{energy2}).
It is amenable to a complete analysis of 
the possible solutions for magnetic structures.
The boundaries of their existence in parameter space
could be calculated in all detail and 
a clear physical picture of 
the formation and evolution 
of these magnetic states is achieved.
Due to the unique combination of those interactions
inducing weak ferromagnetism
and those stabilizing modulated chiral states, 
a rich variety of new modulated and localized structures 
was found to exist in this class of magnetic crystals.
In these inhomogeneous states chiral rotation of 
the staggered magnetization is always accompanied 
by oscillations of a weak magnetization component 
in the basal plane (Fig.~\ref{fig4ab}).
The modulated states in these systems can be realized 
as structures with the propagation vector along 
certain in-plane directions (spirals).
We remark that another type of solutions, two-dimensional
modulated phases, so-called vortex lattices 
may also exist.
In noncentrosymmtric {\textit {ferromagnets}}
they are thermodynamically 
stable under applied fields in certain region of the
phase space.\cite{JETP89}
In Ref.~\onlinecite{AFM89}
vortex-lattices in antiferromagnets lacking 
inversion symmetry with $d=0$,
have been studied theoretically.
Nucleation of such vortex
lattices during the transition 
from the spin-flop phase is discussed
in Ref.~\onlinecite{Nikos01}. 
However, it is still unknown whether these vortex lattices
can be thermodynamically stable in antiferromagnetic materials.

We have described one-dimensional localized structures 
(domain walls or kinks) separating 
domains of homogeneous states.
These differ from similar objects found 
in many other classes of magnetic materials 
by oscillations of the local net magnetization 
and the dependence of their energy 
on the sense of rotation for the
staggered magnetization within the wall. 
These peculiarities of their properties 
should be accessible to experimental verification.
Furthermore, two-dimensional localized structures 
with finite sizes (as axisymmetric vortices 
in the antiferromagnetic phase) are possible topological
defects in these systems. 
They are stabilized  only due to
the chiral interactions. 

In this paper we have deliberately avoided 
a detailed investigations of the 
full model (\ref{energy2}). 
Instead, by introducing a 
simplified model, we have described 
the general features of the magnetic properties 
in noncentrosymmetric tetragonal antiferromagnets.
We expect that this phenomenological description 
will provide a guide for further detailed experimental
investigation of the known noncentrosymmetric 
tetragonal antiferromagnets
and for a search of new crystals belonging to this group.

We also briefly indicate here possible further directions of 
the theoretical investigations.
Fourth-order anisotropies are needed 
to describe orientational processes in the basal plane and
the  peculiarities of magnetic properties near the spin-flop field. 
Future theoretical investigation also should include 
the stray field effects responsible for multidomain
states near the first-order phase transitions.
Similar investigations within the general model (\ref{energy2})
pose a much more complex and challenging task. 
This functional can be considered as a generalized version of 
the nonlinear ${\sigma}$-model, one of the basic models
in the theory of nonlinear physics and solitons.
It is related to many other models in condensed matter physics.
\cite{Salomaa87,Sondhi93,PRL01,Pasquier01,Oswald00}
The further development of the theory should 
involve the investigation
of vortices and vortex lattices as done for other
noncentrosymmetric models.\cite{JETP89,AFM89}
Similar multidimensional localized solutions of nonlinear 
field equations are intensely studied in many other
fields of modern physics.\cite{Salomaa87,PRL01,Oswald00}
Nucleation and evolution of such one-dimensional and two dimensional
modulated patterns have deep physical relations to similar
patterns in superconductivity, \cite{Abrikosov} 
liquid crystals,\cite{Oswald00} and other condensed matter systems 
and even in modern cosmological models.\cite{Zurek96}

\begin{acknowledgments}
We thank M.D. Lumsden and J.R. Thompson for
explanations and communication of unpublished results.
A.N.B.\ thanks H.\ Eschrig 
for support and kind hospitality during his stay 
at IFW Dresden.
He acknowledges further support 
by DFG through Sonderforschungsbereich~422.
U.K.R.\ gratefully acknowledges support by
DFG through grant MU 1015/7-1.
\end{acknowledgments}

\end{document}